\documentclass[prodmode,acmcsur]{acmsmall}
\usepackage{url}
\usepackage{graphicx}
\usepackage{multirow}
\usepackage{pifont}
\usepackage{algorithmic}
\usepackage{amssymb}
\usepackage{amsmath}
\usepackage{subfig}
\usepackage{soul}
\usepackage{color}
\usepackage[ruled]{algorithm2e}

\SetAlFnt{\small}
\SetAlCapFnt{\small}
\SetAlCapNameFnt{\small}
\SetAlCapHSkip{0pt}
\IncMargin{-\parindent}

\acmVolume{9}
\acmNumber{4}
\acmArticle{39}
\acmYear{2014}
\acmMonth{3}

\begin{document}
\newcommand{\xmark}{\ding{55}}
% Page heads
\markboth{J. Tang et al.}{A Survey of Signed Network Mining in Social Media}

% Title portion
\title{A Survey of  Signed Network Mining in Social Media}
\author{JILIANG TANG
\affil{Arizona State University}
YI CHANG
\affil{Yahoo Research}
CHARU AGGARWAL
\affil{IBM T.J. Watson Research Center}
HUAN LIU
\affil{Arizona State University}
}

\begin{abstract}
Many real-world relations can be represented by signed networks with
positive and negative links, as a result of which signed network
analysis has attracted increasing attention from multiple
disciplines. With the increasing prevalence of social media
networks, signed network analysis has evolved from developing and
measuring theories to mining tasks. In this article, we present a
review of mining signed networks in the context of  social media and
discuss some promising research directions and new frontiers. We
begin by giving basic concepts and unique properties and principles
of signed networks. Then we classify and review tasks of signed
network mining with representative algorithms. We also delineate
some tasks that have not been extensively studied with formal
definitions and also propose research directions to expand the field
of signed network mining.
\end{abstract}




\acmformat{Jiliang Tang, Yi Chang, Charu Aggarwal, and Huan Liu, 2016. A Survey of Signed Network Mining in Social Media.}

\begin{bottomstuff}
This material is based upon work supported by, or in part by, the
U.S. Army Research Office (ARO) under contract/grant number 025071
and W911NF-15-1-0328, the Office of Naval Research(ONR) under grant
number N000141410095, and the Army Research Laboratory and was
accomplished under Cooperative Agreement Number W911NF-09-2-0053.

Author's addresses: J. Tang is with Computer Science and Engineering, Arizona State University, Tempe, AZ, 85281.  E-mail: {Jiliang.Tang}@asu.edu; H. Liu is with Computer Science and Engineering, Arizona State University, Tempe, AZ, 85281. E-mail: {Huan.Liu}@asu.edu; C. C. Aggarwal is with IBM T.J. Watson Research Center, 1101
Kitchawan Rd, Yorktown, NY 10598. E-mail: charu@us.ibm.com; and Y. Chang is with Yahoo Research, Yahoo! Inc, Sunnyvalue, CA, USA. E-mail: yichang@yahoo-inc.com.
\end{bottomstuff}

\maketitle

\section{Introduction}

In many real-world social systems, relations between two nodes can
be represented as signed networks with positive  and negative links.
In the 1940s, Heider~\citeyear{heider1946attitudes} studied
perception and attitude of individuals and introduced structural
balance theory, which is  an important social theory for signed
networks. In the 1950s, Cartwright and
Harary~\citeyear{cartwright1956structural} further developed the
theory and introduced the notion of balanced signed graph to
characterize forbidden patterns in social networks. With roots in
social psychology, signed network analysis has attracted much
attention from multiple disciplines such as physics and computer
science, and has evolved considerably from both data- and
problem-centric perspectives.

The early work in the field  was mainly based on signed networks
derived from observations in the physical world such as the
international relationships in Europe from 1872 to
1907~\cite{heider1946attitudes}, relationships among  Allied and
Axis powers during World War II~\cite{axelrod1993landscape}, and the
conflict over Bangladesh's separation from Pakistan in
1971~\cite{moore1978international,moore1979structural}. These signed
networks were typically characterized by a small number of nodes
with dense relationships and were notable for their clean structure.
 With the development of social media,
increasing attention has been focused on signed social networks
observed in online worlds. Signed networks in social media represent
relations among online users where positive links indicate
friendships, trust, and like, whereas negative links indicate foes,
distrust, dislike and antagonism. Examples of signed networks in
social media include trust/distrust in
Epinions\footnote{\url{http://www.epinions.com/}}~\cite{massa2005controversial,leskovec2010predicting}
and friends/foes in
Slashdot\footnote{\url{http://slashdot.org/}}~\cite{kunegis2009slashdot}.
Signed networks in social media often have hundreds of thousands of
users and millions of links, and they are usually very sparse and
noisy. Data for signed network analysis has evolved from offline to
social media networks.

Research problems  have evolved together with the evolution of the
nature of available  data sets for signed network analysis. Signed
networks observed in the physical world are often small but dense
and clean. Therefore, early research about signed networks
had mainly focused on developing theories to explain social
phenomenon in signed
networks~\cite{heider1946attitudes,cartwright1956structural}. Later
on, studies were conducted on
measurements~\cite{harary1959measurement,harary1979matrix,harary1980simple,frank1980balance}
and dynamics of social
balance~\cite{antal2005dynamics,radicchi2007social,radicchi2007universality,marvel2011continuous}.
{  The recent availability of large-scale, sparse and noisy social media networks has encouraged increasing attention on
leveraging data mining, machine learning and optimization
techniques~\cite{kunegis2009slashdot,leskovec2010predicting,yang2012friend,chiang2013prediction,Tang-etal14b}.}
Research problems for signed network analysis have evolved from
developing and measuring theories to mining tasks.

This survey mainly focuses on mining tasks for signed networks in
social media. However, it should be pointed out that (a) we will
review theories originating from signed networks in the physical
world for mining signed networks; and (b) we will review
measurements and dynamics of social balance as basis or objectives
in mining signed networks. Note that since nodes represent users in
social networks, we will use the terms ''node'' and ``user''
interchangeably in this article.

\subsection{Mining Signed Networks in Social Media}

The problem of mining unsigned networks in social media (or networks
with only positive links) has been extensively studied for
decades~\cite{knoke2008social,aggarwal2011introduction,zafarani2014social}.
However, mining signed networks requires dedicated methods because
cannot simply use straightforward  extensions of  algorithms and
theories in unsigned networks~\cite{chiang2013prediction}. First,
the existence of negative links in signed networks challenges many
concepts and algorithms for unsigned networks. For example, node
ranking algorithms for unsigned networks such as
PageRank~\cite{page1999pagerank} and
HITS~\cite{kleinberg1999authoritative} require all links to be
positive. Similarly,  spectral clustering algorithms for unsigned
networks cannot, in general, be directly extended to signed
networks~\cite{kunegis2010spectral}, and the concept of structural
hole in unsigned networks is not applicable to signed
networks~\cite{zhang2016trust}. Second, some social theories such as
balance theory~\cite{heider1946attitudes} and status
theory~\cite{leskovec2010signed} are only applicable to signed
networks, while social theories for unsigned networks such as
homophily may not be applicable to signed
networks~\cite{Tang-etal14b}. In addition, the availability of
negative links brings about unprecedented opportunities and
potentials in mining signed networks. First, it is evident from
recent research that negative links have significant added value
over positive links in various analytical tasks. For example, a
small number of negative links can significantly improve positive
link prediction~\cite{guha2004propagation,leskovec2010predicting},
and they can also improve recommendation performance in social
media~\cite{victor2009trust,ma2009learning}. Second, analogous to
mining unsigned networks, we can have similar mining tasks for
signed networks; however, negative links in signed networks make
them applicable to a broader range of applications. For example,
similar tasks for unsigned networks have new definitions for signed
networks such as community detection and link prediction, while new
tasks and applications emerged for only signed networks such as sign
prediction and negative link prediction.

In this article, we present a comprehensive review of current
research findings about mining signed networks and discuss some
tasks that need further investigation. The major motivation of this
article is two-fold:

\begin{itemize}
\item Negative links in signed networks present two unique types of properties -- (1) distinct topological
properties as opposed to positive links; and (2) collective
properties with positive links. These unique properties determine
that the  basic concepts, principles and properties of signed
networks are substantially different from those of unsigned
networks. Therefore an overview of basic concepts, principles and
properties of signed social networks can facilitate a better
understanding of  the challenges, opportunities and necessity of
mining signed networks.
\item The availability of large-scale signed networks from social media has encouraged a large body of literature in mining signed networks.
On the one hand, a classification of typical tasks can promote a
better understanding. On the other hand, the development of tasks of
mining signed social networks is highly imbalanced -- some tasks are
extensively studied, whereas others have not been sufficiently
investigated. For well-studied tasks, an overview will help users
familiarize themselves with the state-of-the-art algorithms; for
insufficiently studied tasks, it is necessary to give formal
definitions with promising research directions that can enrich
current research.
\end{itemize}

The organization and contributions of the article are summarized as
follows:
\begin{itemize}
\item We give an overview of basic concepts, unique principles, and properties of signed networks in Section
2. We discuss approaches to represent signed networks, topological
properties of the negative networks, and collective properties of
positive and negative links with social theories.
\item We classify the mining tasks of signed social networks into node-oriented, link-oriented and application-oriented tasks.
From Section 3 to Section 5, we review well-studied tasks in each
category with representative algorithms and
\item Mining signed networks is in the early stages of development. We discuss some tasks for each category that have not yet received sufficient
attention in the literature. We discuss formal definitions and
promising research directions.
\end{itemize}

{  The readers of this survey are expected to have some
basic understanding of  social network analysis such as adjacency
matrices, reciprocity and clustering coefficient, data mining
techniques such as clustering and classification, machine learning
techniques such as eigen-decomposition, mixture models, matrix
factorization and optimization techniques such as gradient decent
and EM methods.}

\subsection{Related Surveys and Differences}

A few surveys about signed networks analysis exist in the
literature. One of the earliest surveys may be found
in~\cite{taylor1970balance}. This survey gives an overview of
metrics to measure the degree of social balance for given signed
networks. Very recently, Zheng et al.~\cite{zheng2014social}
provides a comprehensive overview of social balance in signed
networks. {  This survey gives an overview about recent metrics to
measure the degree and the dynamics of social balance as well as the
application of social balance in partitioning signed networks.} With
the increasing popularization of signed networks in social media, a
large body of literature has emerged, which leverages machine
learning, data mining and optimization techniques. This survey
provides a comprehensive overview of this emerging area, along with
a discussion of applications and promising research directions.

Compared to signed networks, there are many more surveys about
unsigned network analysis. These surveys cover various topics in
unsigned network analysis including community
detection~\cite{tang2010community}, node
classification~\cite{bhagat2011node}, link
prediction~\cite{liben2007link} and network
evolution~\cite{aggarwal2014evolutionary}. Surveys are also
available about applications of unsigned networks such as data
classification~\cite{sen2008collective},
recommendation~\cite{tang2013social} and information
propagation~\cite{chen2013information}.

\section{Basis of Signed Networks}

The  basic concepts, principles and properties of signed networks
are related to but distinct from those of unsigned networks. In this
section, we review the representations, distinct properties of
negative links, and collective properties of positive and negative
links with social theories.

\subsection{Network Representation}

A signed network $\mathcal{G}$ consists of a set of $N$ nodes
$\mathcal{U} = \{u_1,u_2,\ldots, u_N\}$, a set of positive links
$\mathcal{E}_p$ and a set of negative links $\mathcal{E}_n$. There
are two major ways to represent a signed network $\mathcal{G}$.

As suggested in~\cite{leskovec2010predicting}, positive and negative
links should be viewed as tightly related features in signed
networks. One way is to represent both positive and negative links
into one adjacency matrix ${\bf A} \in \mathbb{R}^{N \times N}$
where ${\bf A}_{ij} = 1$, ${\bf A}_{ij} = -1$ and ${\bf A}_{ij} = 0$
denote positive, negative and missing links from $u_i$ to $u_j$,
respectively.

The independent analyses of the different networks in signed
networks reveal distinct types of properties and it is important to
consider these distinct topological properties in
modeling~\cite{szell2010multirelational}. Therefore we separate a signed network into a network with only positive links and a network with only negative links, and then use two
adjacency matrices to represent these two networks,
respectively. In particular, it uses ${\bf A}^p \in \mathbb{R}^{N
\times N}$ to represent positive links where ${\bf A}_{ij}^p = 1$
and ${\bf A}_{ij}^p = 0$ denote a positive link and a missing link
from $u_i$ to $u_j$. Similarly ${\bf A}_{ij}^n \in \mathbb{R}^{N
\times N}$ is used to represent negative links where ${\bf A}_{ij}^n
= 1$ and ${\bf A}_{ij}^n = 0$ denote a negative link and a missing
link from $u_i$ to $u_j$.

It is easy to convert one representation into the other  with the
following rules: ${\bf A} = {\bf A}^p - {\bf A}^n$, and ${\bf A}^p =
\frac{|{\bf A}| +{\bf A}}{2}$ and ${\bf A}^n = \frac{|{\bf A}| -{\bf
A}}{2}$ {  where $|{\bf A}|$ is the component-wise absolute value of ${\bf A}$.}

\subsection{Properties of Negative Networks}

There are some well known properties of positive links such as
power-law degree distributions, high clustering coefficient, high
reciprocity, transitivity and strong correlation with similarity.
However, we cannot easily extend  these properties of positive links
to negative links. In this subsection, we will review important
properties of negative links in social media, which are analogous to
those of positive links.

\paragraph{Power-law distributions} It is well known that the distributions of incoming or outgoing positive links
for users usually follow power-law distributions -- a few users with
large degrees while most users have small degrees.
In~\cite{Tang-etal14b}, incoming or outgoing negative links for each
user are calculated and there are two important findings -- (a) in a
signed network, positive links are denser than negative links and
there are many users without any incoming and outgoing negative
links; and (b) for users with negative links, the degree distributions
also follow power-law distributions -- a few users have a large
number of negative links, while most users have  few negative links.

\paragraph{Clustering coefficient} Nodes in networks with positive links are often easy to cluster.
This property is often reflected by their high clustering
coefficients (CC). High values of CC are expected because of the
inherently cohesive nature of positive
links~\cite{coleman1988social}. However, the values of clustering
coefficients for negative links are significantly lower than these
for positive links. This  suggests that many useful properties such
as triadic closure cannot be applied to negative
links~\cite{szell2010multirelational}.

\paragraph{Reciprocity} Positive links show high reciprocity.
Networks with positive links are strongly reciprocal, which
indicates that pairs of nodes tend to form bi-directional
connections, whereas networks with negative links show significantly
lower reciprocity. Asymmetry in negative links is confirmed in the
correlations between the in- and out-degrees of nodes. In- and
out-degrees of positive links are almost balanced, while negative
links show an obvious suppression of such
reciprocity~\cite{szell2010multirelational}.

\paragraph{Transitivity} Positive links show strong transitivity,
which can be explained as ``friends' friends are friends''. {
 The authors of~\cite{Tang-etal14b} examined the
transitivity of negative links on two social media signed networks
Epinions and Slashdot and found that negative links may be not
transitive since they observed both ``enemies' enemies are friends``
and ``enemies' enemies are enemies''.}

\paragraph{Correlation with similarity} Positive links have strong correlations with similarity, which can be explained by two important
social theories, i.e., homophily~\cite{mcpherson2001birds} and
social influence~\cite{marsden1993network}. Homophily suggests that
users are likely to connect to other similar users, while social
influence indicates that users' behaviors are likely to be
influenced by their friends. { Via analyzing two real-world signed
social networks Epinions and Slashdot, the authors
in~\cite{Tang-etal14b} found that users are likely to be more
similar to users with negative links than those without any links,
while users with positive links are likely to be more similar than
those with negative links}. These observations suggest that negative
links in signed social networks may denote neither similarity nor
dissimilarity.
\\
\\
In addition, a recent work conducted a comprehensive signed link
analysis~\cite{beigi2016signed} and found -- (1) users with positive
(negative) emotions are likely to establish positive (negative)
links; (2) users are likely to like their friends' friends and
dislike their friends' foes; and (3) users with higher optimism
(pessimism) are more likely to create positive (negative) links.

\subsection{Collective Properties of Positive and Negative Links}

As shown in the previous subsection, distinct properties are
observed for positive and negative links. When we consider positive
and negative links together, they present collective properties,
which can be explained by two important social theories in signed
networks, i.e., balance theory~\cite{heider1946attitudes} and status
theory~\cite{guha2004propagation,leskovec2010signed}. Next we
present these collective properties by introducing these two social
theories, which have been proven to be very helpful in mining signed
social
networks~\cite{leskovec2010signed,yang2012friend,zheng2014social,kunegis2014applications}.
For example, the signed clustering coefficient and the relative
signed clustering coefficient~\cite{kunegis2009slashdot} are defined
based on the intuition ``the enemy of my enemy is my friend''
implied by balance theory. {Note that balance theory is
developed for undirected signed social networks, whereas status
theory is developed for directed signed social networks}

\subsubsection{Balance Theory}

Balance theory is originally introduced in~\cite{heider1946attitudes} at the
individual level and generalized by Cartwright and
Harary~\cite{cartwright1956structural} in the graph-theoretical
formation at the group level. When the signed network is not
restricted to be complete, the network is balanced if all its cycles
have an even number of negative links. Using this definition, it is
proven in~\cite{harary1953notion} that ``a signed graph is balanced
if and only if nodes can be separated into two mutually exclusive
subsets such that each positive link joins two nodes of the same
subset and each negative link joins nodes from different subsets.''
It is difficult to represent real-world signed networks by balanced
structure.  Therefore, Davis~\citeyear{davis1967clustering} introduced
the notion of a {\em clusterizable graph} -- a signed graph is
clusterizable if there exists a partition of the nodes such that
nodes with positive links are in the same subset and nodes with
negative links are between different subsets.

Later on, researchers have proposed some important metrics to
measure the degree of balance of signed networks. {As mentioned
above, the concept of balance has evolved and been generalized.
Hence, these metrics can be categorized according to their adopted
definitions of balance. Some metrics use the definition of balance
by Cartwright and Harary~\cite{cartwright1956structural} hence they
measure the number of balanced or unbalanced cycles. The ratio of
balanced circles among all possible circles was calculate by using
the adjacency matrix ${\bf A}$~\cite{cartwright1966number}, which
was modified to consider the length of cycles
in~\cite{henley1969goodness}. The time complexity of these metrics
is  $O(n^3)$, which is infeasible for large real-world signed
networks. Terzi and Winkler proposed an efficient spectral algorithm
to estimate the degree of balance for large signed
networks~\cite{terzi2011spectral}.  The definition of balance by
Davis~\citeyear{davis1967clustering} established the correlation
between balance and clustering -- clustering is partition of the
nodes of a given signed network into $k$ clusters, such that each
pair of nodes in the same cluster has a positive link and a negative
link exists between each pair of nodes from different clusters.
Therefore the metrics based on the definition by
Davis~\citeyear{davis1967clustering} measure the number of
disagreements -- the number of negative links inside clusters and
the number of positive links between
clusters~\cite{bansal2004correlation,facchetti2011computing,zheng2014social}.
Actually these metrics later on became criteria to partition signed
networks into clusters (or communities) such as approximation
algorithms were developed for minimizing disagreements by
identifying the optimal number of clusters
in~\cite{bansal2004correlation}. More details about these clustering
algorithms will be discussed in Section~\ref{sec:clustering}. }

Balance theory generally implies that ``the friend of my friend is
my friend'' and ``the enemy of my enemy is my
friend''~\cite{heider1946attitudes}.  Let $s_{ij}$ represent the sign of the link between
the $i$-th node and the $j$-th node where $s_{ij} = 1$ and $s_{ij} =
-1$ denote a positive link and a negative link are observed between
$u_i$ and $u_j$. Balance theory suggests that a triad $\langle u_i,
u_j, u_k \rangle$ is balanced if -- (1) $s_{ij} = 1$ and $s_{jk}=1$,
then $s_{ik} = 1$ ; or (2) $s_{ij}=-1$ and $s_{jk}=-1$, then $s_{ik}
= 1$.

\begin{figure}
    \begin{center}
      \includegraphics[scale=0.45]{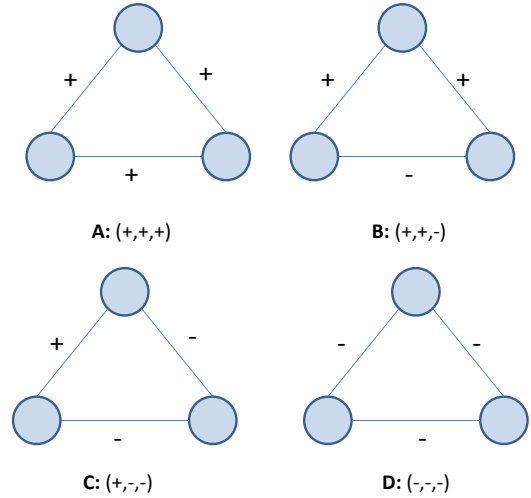}
     \end{center}
\caption{An Illustration of Balance Theory. As suggested by balance theory, triads {\bf A} and {\bf
C} are balanced; while triads {\bf B} and {\bf D} are imbalanced.}
\label{fig:BT}
\end{figure}

For a triad, four possible sign combinations exist as demonstrated
in Figure~\ref{fig:BT}. Among these four combinations, {\bf A} and
{\bf C} are balanced. The way to measure the balance of signed
networks in social media is to examine all these triads and then to
compute the ratio of {\bf A} and {\bf C} over {\bf A}, {\bf B}, {\bf
C} and {\bf D}. Existing work reported that triads in signed
networks in social media are highly balanced. For example, Leskovec
et al.~\cite{leskovec2010predicting} found that the ratios of
balanced triads of signed networks in Epinions, Slashdot and
Wikipedia are $0.941$, $0.912$, and $0.909$, respectively, and more
than $90\%$ of triads are balanced in other social media
datasets~\cite{yang2012friend}. Furthermore, the ratio of balanced
triads increases while that of unbalanced triads decreases over
time~\cite{szell2010multirelational}.

\subsubsection{Status Theory}

While balance theory is  naturally defined  for undirected networks,
status theory~\cite{guha2004propagation,leskovec2010signed} is
relevant for directed networks. Social status can be represented in
a variety of ways, such as the  rankings of nodes in social
networks, and it represents the  prestige of nodes. In its most
basic form, status theory suggests that $u_i$ has a higher status
than $u_j$ if there is a positive link from $u_j$ to $u_i$ or a
negative link from $u_i$ to $u_j$.

\begin{figure}
    \begin{center}
     \includegraphics[scale=0.6]{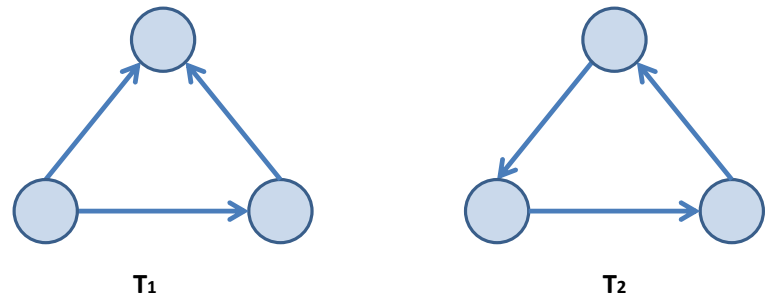}
     \end{center}
\caption{Possible Triads in Directed Social Networks.}
\label{fig:tST}
\end{figure}

\begin{figure}
    \begin{center}
      \includegraphics[scale=0.5]{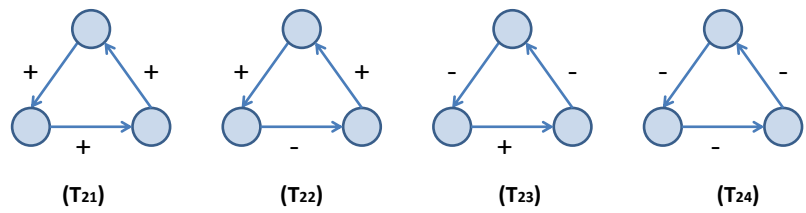}
     \end{center}
\caption{An Illustration of Four Types of Cyclic Signed Triads.}
\label{fig:sAB}
\end{figure}

\begin{figure}
    \begin{center}
      \includegraphics[scale=0.5]{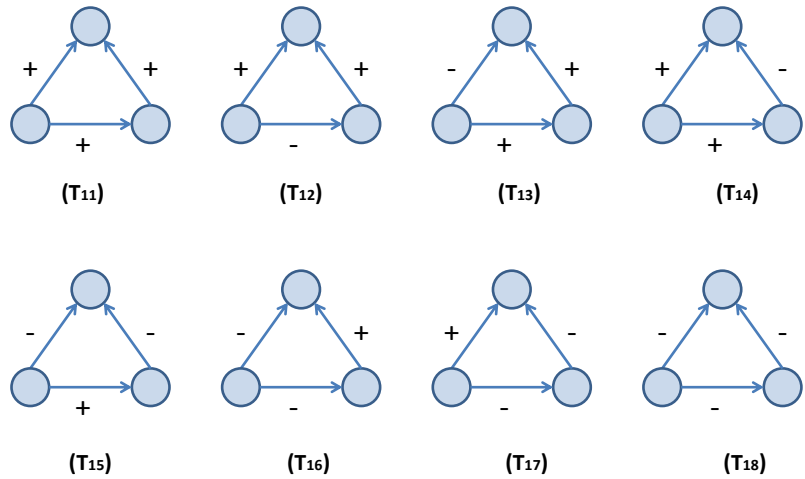}
     \end{center}
\caption{An Illustration of Eight Types of Acyclic Signed Triads.}
\label{fig:sCD}
\end{figure}

As shown in Figure~\ref{fig:tST},  there are two types of triads in
directed networks,  which correspond to acyclic and cyclic triads.
Note that flipping the directions of all the links has no impact on
the type of the cyclic triad.   Since there are four possible sign
combinations, there are four types of cyclic signed triads for $T_2$
as shown in Figure~\ref{fig:sAB}. Each link in an acyclic triad can
be positive or negative and the  signs of links in an acyclic triad
are not exchangeable; hence, there are eight types of acyclic signed
triads as depicted in Figure~\ref{fig:sCD}. Overall, there are 12
types of triads in directed signed networks.

A popular approach to examine whether a given triad satisfies status
theory or not is as follows. We reverse the directions of all
negative links and flip their signs to positive. If the resulting
triad is acyclic, then the triad satisfies status theory.  It is
easy to verify that  (1) for a negative link $u_i \xrightarrow{-}
u_j$,  reversing its direction and flipping its sign simultaneously
lead to a positive link $u_j \xrightarrow{+}u_i$, which preserves
the status order of $u_i$ and $u_j$ according to status theory; and
(2) for a positive and cyclic triad $u_i \xrightarrow{+} u_j
\xrightarrow{+} u_k \xrightarrow{+} u_i$,  their statuses should
satisfy $u_i > u_j > u_k > u_i$ according to status theory, which
leads to a logical contradiction $u_i > u_i $. Following the
aforementioned approach, we find that 8 of the 12 types of triads in
signed networks satisfy status theory as shown in the first row of
Table~\ref{tab:BST}. Similar to the approach for testing the balance
of signed networks, we examine all 12 triads and then  calculate the
ratio of triads satisfying status theory. Examinations on signed
networks in typical social media suggest that more than $90\%$ of
triads satisfy status theory~\cite{leskovec2010signed}.

\begin{table}
\caption{Balance Theory vs. Status Theory.}
\centering
\begin{tabular}{|c|c|c|c|c|c|c|c|c||c|c|c|c|}
\hline
 & $T_{11}$ & $T_{12}$ & $T_{13}$ & $T_{14}$ & $T_{15}$ & $T_{16}$ & $T_{17}$ & $T_{18}$ & $T_{21}$ & $T_{22}$ & $T_{23}$ & $T_{24}$ \\\hline
Status Theory &\checkmark &\checkmark &\xmark &\checkmark &\checkmark&\checkmark&\xmark &\checkmark &\xmark&\checkmark&\checkmark&\xmark  \\\hline
Balance Theory&\checkmark &\xmark &\xmark &\xmark &\checkmark&\checkmark&\checkmark &\xmark&\checkmark&\xmark&\checkmark&\xmark \\
  \hline
\end{tabular}
\label{tab:BST}
\end{table}

As shown in Table~\ref{tab:BST}, status theory and balance theory do
not always agree with one another. Note that we apply balance theory
to directed signed networks by ignoring the directions of links.
Some triads satisfy both theories such as the triad $T_{11}$. Some
satisfy status theory but not  balance theory such as the triad
$T_{12}$. Some satisfy balance theory but not status theory such as
the triad $T_{21}$. Others do not satisfy either such as the triad
$T_{24}$.

\subsection{Popular Data Sets for Benchmarking}

In this subsection, we discuss some social media data sets widely
used for benchmarking analytical algorithms in the  signed network
setting.

Epinions is a product review site. Users can create both positive
(trust) and negative (distrust) links to other users. They can write
reviews for various products with rating scores from $1$ to $5$.
Other users can rate the helpfulness of reviews. There are several
variants of datasets from Epinions publicly
available~\cite{massa2005controversial,leskovec2010predicting,yang2012friend,Tang-etal15a}.
Statistics of two representative sets are illustrated  in
Table~\ref{tab:datasets}. ``Epinions'' is from Stanford large
network dataset
collection~\footnote{https://snap.stanford.edu/data/} where only
signed networks among users are available. In addition to signed
networks, ``eEpinion''~\cite{Tang-etal15a}  also provides item
ratings, review content, helpfulness ratings and categories of
items.  It also includes timestamps when links are established and
ratings are created.

Slashdot is a technology news platform in which users can create
friend (positive) and foe (negative) links to other users. They can
also post news articles. Other users may annotate these articles
with their comments. There also various variants of datasets from
Slashdot~\cite{kunegis2009slashdot,leskovec2010predicting,Tang-etal15a}
and two of them are demonstrated in Table~\ref{tab:datasets}.
``Slashdot'' is from Stanford large network dataset collection with
only signed networks among users, while the more detailed
``eSlashdot''~\cite{Tang-etal15a} provides signed networks, comments
on articles, user tags and groups in which users participate.

\begin{table}
\caption{Statistics of Representative Signed Networks in Social Media.}
\centering
\begin{tabular}{|c|c|c|c|c|} \hline
                                                     &Epinions     &Slashdot    &eEpinions  & eSlashdot  \\ \hline
\# of Users                                    &119,217      &82,144       &23,280       &14,799     \\ \hline
\# of Links                                     &841,200      &549,202     &332,214     &232,471  \\ \hline
Positive Link Percentage              &85.0\%        &77.4\%       &87.7\%      &81.5 \%    \\ \hline
Negative Link Percentage            &15.0\%        &22.6\%        &12.3\%      &18.5 \%  \\ \hline
\end{tabular}
\label{tab:datasets}
\end{table}

\subsection{Tasks of Mining Signed Networks}

\begin{figure*}
    \begin{center}
   \includegraphics[scale=0.6]{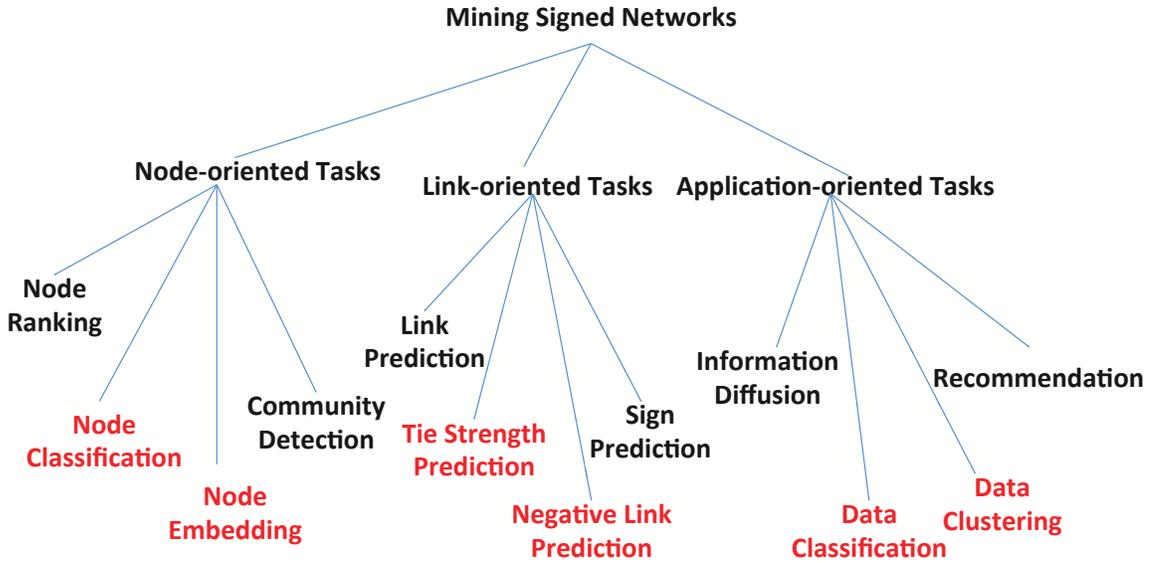}
    \end{center}
\caption{An Overview of Tasks of Mining Signed Networks in Social
Media. Tasks highlighted in red have not been extensively studied.}
\label{fig:over}
\end{figure*}

There are similar tasks for mining unsigned and signed networks.
However, the availability of negative links in signed networks
determines that similar mining tasks for unsigned networks may have
new definitions for signed networks and there may be new tasks
specific to signed networks. We category the tasks of mining signed
networks as tasks that focus on nodes, links and applications, i.e., node-oriented, link-oriented and application-oriented
tasks as shown in Figure~\ref{fig:over}. Although a large body of
work has emerged in recent years for mining signed social networks,
the development of tasks in each category is highly imbalanced. Some
of them are well studied, whereas others need further investigation.
These tasks are highlighted in  red in Figure~\ref{fig:over}. In the
following sections,  we give an overview of representative
algorithms for well-studied tasks and also provide a detailed
discussion of  important and emerging tasks. Where needed, promising
research directions are also highlighted.  The  notations used in
this article are summarized in Table~\ref{tab:notations}.
 {  Any algorithms for directed signed networks are applicable to undirected signed networks
 by considering an undirected link as two bidirectional links. Hence, in the following
 sections, it can be assumed by default that
 an algorithm can be applied to both directed and undirected signed networks.}

\begin{table}
\caption{Notation and Definitions.}
\centering
\begin{tabular}{|c|l|}
\hline
  Notations      & Descriptions               \\ \hline
  $N$            & Number of Users        \\
  ${\bf A}$      & Adjacency matrix of a signed network \\
  ${\bf A}^p$    & Adjacency matrix of a positive network \\
  ${\bf A}^n$    & Adjacency matrix of a negative network \\
  ${\bf D}^p$    & A diagonal matrix with ${\bf D}_{ii}^p = \sum_j {\bf A}_{ij}^p$\\
  ${\bf D}^n$    & A diagonal matrix with ${\bf D}_{ii}^n = \sum_j {\bf A}_{ij}^n$\\
  $I_i^+$        & The set of nodes that create positive links to $u_i$ \\
  $|I_i^+|$       & Indegree of positive links of $u_i$ \\
  $I_i^-$        & The set of nodes that create negative links to $u_i$ \\
  $|I_i^-|$        & Indegree of negative links of $u_i$ \\
  $I_i$            & $I_i =  I_i^+ \cup I_i^-$ \\
  $| I_i |$         & $| I_i | = |I_i^+|  + |I_i^-|$ \\
  $O_i^+$        & The set of users that $u_i$ creates positive links to \\
  $|O_i^+|$        & Outdegree of positive links of $u_i$ \\
  $O_i^-$        & The set of users that $u_i$ creates negative links to \\
  $|O_i^-|$        & Outdegree of negative links of $u_i$ \\
  $O_i$          & $O_i =  O_i^+ \cup O_i^-$ \\
  $|O_i|$          & $|O_i| =  |O_i^+| + |O_i^-|$ \\
  $d_i^+$        & $d_i^+ = |I_i^+| + |O_i^+|$  \\
  $d_i^-$        & $d_i^- = |I_i^-| + |O_i^-|$ \\
  $\mathcal{L}^p$ & Laplacian matrix for a positive network \\
  $\mathcal{L}^n$ & Laplacian matrix for a negative network \\
  $\mathcal{L}$   & Laplacian matrix for a signed network \\
  $c_i$           & Community of $u_i$ \\
  $s_{ij}$        & Sign of the link from $u_i$ to $u_j$ \\
  $m$             & Number of links in a signed social network\\
   $m^+$        & Number of positive links in a signed social network \\
  $m^-$         & Number of negative links in a signed social network \\
   ${\bf X}_{ij}$    & the $(i,j)$ entry of the matrix ${\bf X}$ \\
  \hline
\end{tabular}
\label{tab:notations}
\end{table}

\section{Node-oriented Tasks}
As shown in Figure~\ref{fig:over}, important node-oriented tasks
include node ranking, community detection, node classification and
node embedding, among which node ranking and community detection
are extensively studied. On the other hand,   node classification
and node embedding need further investigations. In this section,
we review node ranking and community detection with representative
algorithms.

\subsection{Node Ranking}

The problem of node ranking for signed networks is that of
exploiting the link structure of a network to order or prioritize
the set of nodes within the network by considering both positive and
negative links~\cite{getoor2005link}. Since negative links are
usually not considered, most node ranking algorithms for unsigned
networks cannot deal with negative values
directly~\cite{haveliwala2002topic,cohn2000learning}. A
straightforward solution is to apply node ranking algorithms of
unsigned networks, such as EigenTrust~\cite{kamvar2003eigentrust},
by ignoring negative links or zero the entries corresponding to the
negative links in the matrix representation of the
network~\cite{richardson2003trust}. In other words,   we only
consider the positive network ${\bf A}^p$ while ignoring the impact
from ${\bf A}^n$ in a signed network. This solution cannot
distinguish between negative and missing links since both of them
correspond to a zero entity in the representation matrix. Recent
node ranking algorithms  for signed networks fall into three themes
-- (a)  centrality measurements are used; (b) PageRank-like models
are used~\cite{page1999pagerank}; and (c) HITS-like methods are
used~\cite{kleinberg1999authoritative}.
 Next, we will introduce representative
algorithms for each group.

\subsubsection{Centrality-based algorithms}

Centrality-based algorithms use certain centrality measurements to
rank nodes in signed networks. If a node receives many positive
incoming links, it should have high prestige value, while nodes with
many negative incoming links will have small values of prestige.{  A measure $p_i$ of the status
score of $u_i$ based on the indegree of positive and negative links is
proposed in~\cite{zolfaghar2010mining} as follows:
\begin{align}
p_i = \frac{|I_i^+| - |I_i^-|}{|I_i^+| + |I_i^-|}
\end{align}
\noindent where $|I_i^+|$, and $|I_i^-|$ are the indegree of
positive and negative links of $u_i$, respectively. }A similar
metric is used in~\cite{kunegis2009slashdot} as the subtraction of
indegree of negative links from indegree of positive links, i.e.,
$p_i = |I_i^+| - |I_i^-|$. An eigenvector centrality metric is
proposed in~\cite{bonacich2004calculating} for balanced complete
signed networks. We can divide nodes of a balanced complete signed
network into two communities such that all positive links connect
members of the same community and all negative links connect members
of different communities. {  Thus, positive and negative
scores in the eigenvector that correspond to the largest eigenvalue
of the adjacency matrix ${\bf A}$ reveal not only the clique
structure but also status scores within each
clique~\cite{bonacich2004calculating}.}

\subsubsection{PageRank-based Algorithms}

The original PageRank algorithm expresses the reputation score for the $i$-th node as:
\begin{align}
p_i = \sum_{u_j \in I_i^+} \frac{p_j}{|O^+_j|}
\end{align}
\noindent where $|O^+_j|$ is the outdegree of positive links of
$u_j$. The probability $p_i$ can be computed in an iterative way:
\begin{align}
p_i^{t+1} = \alpha \sum_{u_j \in I_i^+} \frac{p_j^t}{|O^+_j|} + (1-\alpha) \frac{1}{N}
\end{align}
\noindent where the term $(1-\alpha) \frac{1}{N}$ is the restart
component, $N$ the total number of users, and $\alpha$ is a  damping
factor. In  signed networks, mechanisms are also provided  to handle
negative
links~\cite{traag2010exponential,borgs2010novel,chung2013dirichlet}.
Next, we detail three representative algorithms in this
group~\cite{shahriari2014ranking,de2008pagetrust,traag2010exponential}

In~\cite{shahriari2014ranking}, two status scores are calculated by
the original PageRank algorithm for the positive network and the
negative network separately, and the difference of the two provides
the final result. Therefore, this  algorithm considers a signed
network as two separate networks and completely ignores the
interactions between positive and negative links. Furthermore, this approach does not have natural interpretations in terms of the reputation scores of nodes. In~\cite{wu2016troll}, the Troll-Trust model is proposed that has a clear physical interpretation. An exponential
node ranking algorithm based on discrete choice theory is proposed
in~\cite{traag2010exponential}.   When the observed reputation is
$k_i = \sum_{u_j \in I_i} {\bf A}_{ji} p_j$, the probability of
$u_i$ with the highest real reputation according to discrete choice
theory is:
\begin{align}
p_i = \frac{exp(k_i / \mu)}{\sum_j exp(k_j / \mu)}
\end{align}
\noindent An iterative approach is used to compute the status scores
as follows:
\begin{align}
{\bf p}^{t+1} = \frac{exp (\frac{1}{\mu} {\bf A}^\top {\bf p}^t) }{\| exp (\frac{1}{\mu} {\bf A}^\top {\bf p}^t) \|_1}
\end{align}
Within a certain range of $\mu$, the aforementioned formulation can
achieve a global solution ${\bf p}^*$ with arbitrary
initializations.

 The work
in~\cite{de2008pagetrust,de2009ranking} uses the intuition that the
random-walk process should be modified to avoid negative links.
Therefore nodes receiving negative connections are visited less.
This is formalized as follows:
\begin{align}
{\bf p}_i^{t+1} = (1 - \hat{ {\bf Q}}_{ii}^t)(\alpha \sum_{u_j \in I^+_i} \frac{{\bf p}_j^t}{|O_j^+|} + (1-\alpha) \frac{1}{N})
\end{align}
\noindent where $\hat{ {\bf Q}}_{ii}^t$ gives the ratio of walkers
that distrust the node they are in. In that manner $(1 - \hat{ {\bf
Q}}_{ii}^t)$ represents the ratio of remaining walkers in $u_i$. {The
distrust matrix $ \hat{ {\bf Q} }$ is calculated as follows}:
\begin{itemize}
\item A random walk according to the original PageRank formulation
is used:
\begin{align}
\hat{{\bf Q}}^{t+1} = {\bf T}^t {\bf Q}^t
\end{align}
\noindent {  where ${\bf T}^t$ is the transition  matrix
whose $(i,j)$th entry ${\bf T}_{ij}^t$ indicates the ratio of
walkers in $u_i$ who were in $u_j$ at time $t$ as follows}:
\begin{align}
{\bf T}_{ij}^t = \frac{\alpha {\bf A}_{ij}^p {\bf p}_j^t / |O_j^+| + (1 - \alpha) \frac{1}{N} }  {\alpha \sum_{u_k \in I^+_i} \big( {\bf p}_k^t /   |O_k^+| + (1 - \alpha) \frac{1}{N} \big) }
\end{align}
\item A walk in $u_i$ automatically adopts negative opinions of $u_i$. In other words,
 he adds the nodes negatively pointed by $u_i$ into his distrust list (${\bf Q}_{ij}^{t+1} = -1$). A walker who distrusts a node leaves
 the graph if ever he visits the node (${\bf Q}_{ij}^{t+1} = 0$). {With the intuition, ${\bf Q}_{ij}^{t+1}$ is updated as follows}:
\begin{align}
{\bf Q}_{ij}^{t+1} =
\left\{
\begin{array}{l}
1~~~~~ \text{if ${\bf A}_{ij} = -1$ }, \\
0~~~~~~\text{if $i = j$}, \\
\hat{{\bf Q}}_{ij}^{t+1}~~~~~~~\text{otherwise}
\end{array}
\right.
\end{align}
\end{itemize}

\subsubsection{HITS-based Algorithms}

The original HITS algorithm~\cite{kleinberg1999authoritative} calculates a hub score $h_i$ and an authority score $a_i$ for each node $u_i$ as
\begin{align}
h_i = \sum_{j \in I_i^+} a_j;~~~ a_i = \sum_{j \in O_i^+} h_j
\end{align}

HITS-based algorithms provide components to handle negative links
based on the original HITS algorithm.
In~\cite{shahriari2014ranking}, two strategies are proposed. The
first applies the original HITS algorithm separately on the positive
and negative networks as follows:
\begin{align}
\left\{
\begin{array}{l}
h_i^+ = \sum_{j \in I_i^+} a_j^+;~~~ a_i^+ = \sum_{j \in O_i^+} h_j^+ \\
h_i^- = \sum_{j \in I_i^+} a_j^-;~~~ a_i^- = \sum_{j \in O_i^-} h_j^-
\end{array}
\right.
\end{align}
\noindent Then, the final hub and authority scores are computed as
follows:
\begin{align}
a_i = a_i^+ - a_i^-;~~~ h_i = h_i^+ - h_i^-
\end{align}
\noindent The other way is to incorporate the  signs  directly as
follows:
\begin{align}
\left\{
\begin{array}{l}
h_i = \frac{\sum_{j \in I^+} a_j - \sum_{j \in I_i^-} a_j}{\sum_{j \in I_i^+} a_j + \sum_{j \in I_i^-} a_j} \\
a_i = \frac{\sum_{j \in O_i^+} h_j - \sum_{j \in O_i^-} h_j}{\sum_{j \in O_i^+} h_j + \sum_{j \in O_i^-} h_j}
\end{array}
\right.
\end{align}

Instead of hub and authority scores in HITS, the concepts of {\em
bias} and {\em deserve} are introduced in~\cite{mishra2011finding}.
Here, bias (or trustworthiness) of a link reflects the expected
weight of an outgoing connection and deserve (or prestige) of a link
reflects the expected weight of an incoming connection from an
unbiased node. Similar to HITS, the deserve score $DES_i$ for $u_i$
is the aggregation of all unbiased votes from her incoming
connections as:
\begin{align}
DES_i^{t+1} = \frac{1}{|I_i|} \sum_j {\bf A}_{ji}(1 - X_{ji}^t)
\end{align}
\noindent where $X_{ji}$ indicates the influence that bias of $u_j$ has on its outgoing link to $u_i$
\begin{align}
X_{ji} = \max\{0,BIAS_j* {\bf A}_{ji}\}
\end{align}
\noindent while the bias score $BIAS_i$ for $u_i$ is the aggregation of voting biases of her outgoing connections as:
\begin{align}
BIAS_i^{t+1} = \frac{1}{2*|O_i|} \sum_{u_j \in O_i} (A_{ji} - DES_j^t)
\end{align}

\subsection{Community Detection in Signed Networks}

The existence of negative links in signed networks makes the
definition of community detection in signed networks substantially
different from that in unsigned networks. In unsigned networks,
community detection identifies groups of densely connected
nodes~\cite{tang2010community,papadopoulos2012community,ailon2013breaking}. In signed
networks,  groups of users are identified, where users are densely
connected by positive links within the group and negative links
between groups.  Based on the underlying methodology,
clustering-based, modularity-based, mixture-model-based and dynamic-model-based methods are used.
Next we will give basic concepts for each group with representative
algorithms.

\subsubsection{Clustering-based Algorithms}
\label{sec:clustering} Clustering-based algorithms transform a graph
vertex clustering problem to one that can be addressed by
traditional data clustering methods. If we consider a positive link
or a negative link indicates whether two nodes are similar or
different, community detection in signed networks is boiled down to
the correlation clustering problem~\cite{bansal2004correlation}.
Bansal et. al. proved NP hardness of the correlation clustering
problem and gave constant-factor approximation algorithms for the
special case in {which the network is complete and undirected,  and
every edge has weight $+1$ or $-1$~\cite{bansal2004correlation}. }A
two phase clustering re-clustering algorithm is introduced
in~\cite{sharma2009community} -- (1) the first phase is based on
Breadth First Search algorithm which forms clusters on the basis of
the positive links only; and (2) the second phase is to reclassify
the nodes with negative links on the basis of the participation
level of the nodes having the negative links. In addition, there are
two groups of clustering algorithms for community detection.  One is
based on k-balanced social theory and the other is based on spectral
clustering. { Note that algorithms based on spectral clustering are
designed for undirected signed networks.}

{  Algorithms based on k-balanced social theory aim to
find $k$ clusters with minimal positive  links between clusters and
minimal negative links inside clusters}.
In~\cite{doreian1996partitioning}, the objective function of
clustering algorithms is defined as $E = \alpha N_n +
(1-\alpha)N_p$, where $N_n$ is the number of negative links within
clusters and $N_p$ the number of positive links between clusters.
The proposed algorithm
in~\cite{doreian1996partitioning,hassan2012detecting} first assigns
the nodes to $k$ clusters randomly, and then optimizes the above
objective function through reallocating the nodes. An alternative
approach is to leverage  simulated annealing  to optimize the
objective function
$E$~\cite{traag2009community,bogdanov2010towards}.

{  One spectral clustering technique is introduced
in~\cite{kunegis2010spectral}. For a signed network ${\bf A}$, it
first defines the signed Laplacian matrix~\cite{hou2005bounds} as
follows:}
\begin{align}
\mathcal{L} = {\bf D} - {\bf A},~~{\bf D}_{ii} = \sum_j |{\bf A}_{ij}|
\label{eq:sl}
\end{align}
\noindent Similar to the Laplacian matrix for unsigned networks, it
can be proven that the signed Laplacian matrix $\mathcal{L}$ is
often positive-semidefinite but it is positive-definite if and only
if  the network is unbalanced. Spectral clustering algorithms on the
signed Laplacian matrix can detect clusters of nodes within which
there are only positive links. The Laplacian matrix in
Eq.~(\ref{eq:sl} tends to separate pairs with negative links rather
than to force pairs with positive links closer. Hence a balanced
normalized signed Laplacian matrix is proposed
in~\cite{zhengspectral} as:
\begin{align}
\mathcal{L} = ({\bf D}^p - {\bf A}^p  + {\bf A}^n)
\label{eq:bsl}
\end{align}
\noindent

Another spectral clustering technique is balanced normalized
cut~\cite{chiang2012scalable}. The objective of a positive ratio cut
is to minimize the number of positive links between communities:
\begin{align}
\min \sum_{c = 1}^k \frac{x_c^\top \mathcal{L}^p x_c }{x_c^\top x_c}
\label{eq:lp}
\end{align}
\noindent where $\{x_c\}_{c=1}^k$ are the community indicator
vectors, and $\mathcal{L}^p$ is the Laplacian matrix of positive
links.  The objective of negative ratio association is to minimize
the number of negative links in each cluster as:
\begin{align}
\min (\sum_{c = 1}^k \frac{x_c^\top {\bf A}^n x_c }{x_c^\top x_c})
\label{eq:an}
\end{align}
{The balance normalized cut is to minimize the positive ratio cut and negative ratio association simultaneously as:
\begin{align}
\min (\sum_{c = 1}^k \frac{x_c^\top ({\bf D}^p - {\bf A}) x_c }{x_c^\top x_c})
\label{eq:dn}
\end{align}
\noindent where the matrix of ${\bf D}^p - {\bf A}$ in balanced
normalized cut is identical to the balanced normalized signed
Laplacian matrix in Eq.~(\ref{eq:bsl}).}

{ We can obtain $\{x_1,x_2,\ldots,x_k\}$ by solving the optimization problems in Eqs~(\ref{eq:lp}), (\ref{eq:an}) or (\ref{eq:dn}).
To generate $k$ clusters, we can round $\{x_1,x_2,\ldots,x_k\}$ to a valid indicator set~\cite{chiang2012scalable} -- we
consider $(x_1(i), x_2(i), \ldots, x_k(i))$ as a $k$-dimensional vector representation of user $i (i\in\{1,2,\ldots,n\} )$ and
then perform $k$-means on these $n$ vectors}

\subsubsection{Modularity-based Algorithms}

These algorithms are to detect communities by optimizing modularity
or its variants for signed networks~\cite{li2014comparative}. The
original modularity~\cite{newman2004finding} is developed for
unsigned networks and it measures how far the real positive
connections deviates from the expected random connections, which is
formally defined as follows: {
\begin{align}
Q^+ = \frac{1}{2m^+} \sum_{ij}( {\bf A}_{ij}^p  - \frac{d_i^+ d_j^+}{2m^+}) \delta(i,j)
\end{align}}
\noindent where $\delta(c_i,c_j)$ is the Kronecker delta function which is 1 if $u_i$ and $u_j$ are in the same community, and 0 otherwise. In~\cite{gomez2009analysis}, modularity of networks with only negative links $Q^-$ is defined in a similar as $Q^+$: {
\begin{align}
Q^- = \frac{1}{2m^-} \sum_{ij}( {\bf A}_{ij}^n  - \frac{d_i^- d_j^-}{2m^-}) \delta(i,j)
 \end{align}}
 \noindent Modularity for signed network $Q$ should balance the tendency of users with positive links to form communities and that of users with negative links to destroy them and the mathematical expression of $Q$ is:
\begin{align}
Q = \frac{2 m^+}{ 2 m^+ + 2 m^-} Q^+ -  \frac{2 m^-}{ 2 m^+ + 2 m^-} Q^-
\label{eq:smodu}
\end{align}
Eq.~(\ref{eq:smodu}) can be rewritten as:
\begin{align}
Q = \frac{1}{2m} \sum_{ij}( {\bf A}_{ij} + \frac{d_i^- d_j^-}{2m^-} - \frac{d_i^+ d_j^+}{2m^+}) \delta(i,j)
\end{align}

The definition of $Q$ in Eq.~(\ref{eq:smodu}) has three properties~\cite{li2014comparative} -- (1) $Q$ boils down to $Q^+$ if no negative link exists; (2) $Q = 0$ if all nodes are assigned to the same community; and (3) $Q$ is anti-symmetric in weighted signed networks.
 {Based on $Q$ in Eq.~(\ref{eq:smodu}), several variants of modularity are developed such as modularity density~\cite{li2014comparative} and frustration~\cite{anchuri2012communities}. Community structure can be obtained by either minimizing frustration~\cite{anchuri2012communities}
or maximizing modularity}, both of which have  been proven to
be  a general eigenvector problem~\cite{anchuri2012communities}.
In~\cite{amelio2013community}, a community detection framework
SN-MOGA is proposed by using non-dominated sorting
genetic~\cite{srinivas1994muiltiobjective,pizzuti2009multi} to
minimize frustration and maximize signed modularity simultaneously.

\subsubsection{Mixture-model-based Algorithms}

Mixture-model-based algorithms generate the division of the network into communities based on generative graphical models~\cite{chen2013overlapping}. In general, there are two advantages of mixture-model-based algorithms. First they provide soft-partition solutions in signed networks. Second, they provide soft-memberships which indicate the strength of a node belonging to a community. These two advantages determine that they can identify overlapping communities. Stochastic block-based models and probabilistic mixture-based models are two types of mixture models widely adopted for community detection in signed networks.  Stochastic block-based models generate a network from a node perspective where each node is assigned to a block or community and links are independently generated for pairs of nodes. In~\cite{jiang2015stochastic}, a generalized stochastic model, i.e., signed stochastic block model (SSBM), is proposed to identify communities for signed networks where nodes within a community  are more similar in terms of positive and negative connection patterns than those from other communities. SSBM represents the memberships of each node as hidden variables and uses two matrices to explicitly characterize positive and negative links among groups, respectively.  While probabilistic mixture-based models generate a network from the link perspective~\cite{shen2013community}. In~\cite{chen2013overlapping} , a signed probabilistic mixture (SPM) model is proposed to detect overlapping communities in undirected signed networks. A link from $u_i$ to $u_j$ is generated by SPM as follows:
\begin{itemize}
\item If the link from $u_i$ to $u_j$ is positive, i.e., ${\bf A}_{ij} > 0$:
\begin{enumerate}
\item Choose a community $c$ for the link with probability $W_{cc}$
\item Select $u_i$ from $c$ with probability $\theta_{ci}$
\item Select $u_j$ from $c$ with probability $\theta_{cj}$
\end{enumerate}
\item If the link from $u_i$ to $u_j$ is negative, i.e., ${\bf A}_{ij} < 0$:
\begin{enumerate}
\item Choose two different communities $c$ and $s$ for the link with probability $W_{cs (c \neq s)}$
\item Select $u_i$ from  $c$ with probability $\theta_{ci}$
\item Select $u_j$ from  $s$ with probability $\theta_{sj}$
\end{enumerate}
\end{itemize}
Overall, the probability of the link from $u_i$ to $u_j$ can be rewritten as:
\begin{align}
P({\bf A}_{ij} | W, \theta) = (\sum_{cc} W_{cc} \theta_{ci} \theta_{cj})^{{\bf A}_{ij}^p}  (\sum_{cs (c \neq s)} W_{cs} \theta_{ci} \theta_{sj})^{{\bf A}_{ij}^n}
\end{align}

\subsubsection{Dynamic-model-based Algorithms}

Dynamic-model-based algorithms consider a dynamic process taking
place on the network, which reveals its communities. {
One type of algorithm in this group is based on discrete-time and
continuous-time dynamic models of social balance}, and a review of
these algorithms can be found in~\cite{zheng2014social}.
{ A framework based on agent-based random walk model  is
proposed in~\cite{yang2007community} to extract communities for
signed networks. Generally, links are much denser within a community
than between communities. The intuition behind this framework is
that an agent, starting from any node, should have higher chances to
stay in the same community than to go to a different community after
a number of walks.  The framework has two advantages -- (1) it is
very efficient with linear time complexity in terms of the number of
nodes; and (2) it considers both the density of links and signs,
which provides a unified framework for community detection for
unsigned and signed networks. Some additional steps are added
by~\cite{kong2011improvement} to further advance the framework such
as introducing a method to detect random walk steps automatically. }

%With
%this intuition, the proposed framework identifies communities by
%examining these transition probabilities including two phases -- (1)
%the first phase performs random walks to transform the adjacency matrix
%to transition probabilities and then sorts them by rows, and (2) the
%second phase divides the transformed matrix into two block matrices,
%which can be used to identify two sub-graphs. One corresponds to a
%identified community, and the other is recursively processed by the two phases.

\subsection{Promising Directions for Node-oriented Tasks}
In this subsection, we discuss two node-oriented tasks including
node classification and node embedding, which need further
investigations to help us gain a better understanding of nodes in
signed networks.

\subsubsection{Node Classification in Signed Networks}

User information such as demographic values, interest  beliefs or
other characteristics plays an important role in helping social
media sites provide better services for their users such as
recommendations and content filtering. However, most social media
users do not share too much of their
information~\cite{zheleva2009join}. For example, more than 90\% of
users in Facebook do no t reveal their political
views~\cite{abbasi2014scalable}. One way of bridging this knowledge
gap is to infer missing user information by leveraging the
pervasively available network structures in social media. An example
of such inference is that of node classification in social networks.
The node classification problem has been extensively studied in the
literature~\cite{getoor2005link}. The vast majority of these
algorithms have focused on unsigned social networks (or social
networks with only positive links)~\cite{sen2008collective}.
Evidence from recent achievements in signed networks suggests that
negative links may be also potentially helpful in the task of node
classification.

Let $\mathcal{C} = \{c_1,c_2,\ldots,c_m\}$ be the set of $m$ class
labels. Assume that $\mathcal{U}^L = \{u_1,u_2,\ldots,u_n\}$ is the
set of $n$ labeled users where $n < N$ and $\mathcal{U}^U =
\mathcal{U} \backslash \mathcal{U}^L$ is the set of $N-n$ unlabeled
users. We use ${\bf Y} \in \mathbb{R}^{n \times m}$ to denote the
label indicator matrix for $\mathcal{U}^L$ where ${\bf Y}_{ik} = 1$
if $u_i$ is labeled as $c_k$, ${\bf Y}_{ik} = 0$ otherwise. With
above notations and definitions, the problem of user classification
in a signed social network can be formally stated as
follows: {\it Given a signed social network
$\mathcal{G}$ with ${\bf A}^p$ and ${\bf A}^n$, and labels ${\bf Y}$
for some users $\mathcal{U}^L$, user classification in a signed
social network aims to infer labels for $\mathcal{U}^U$ by
leveraging ${\bf A}^p$, ${\bf A}^n$ and ${\bf Y}$.}

There are two possible research directions for node classification
in signed networks. Since node classification has been extensively
studied for unsigned networks, one way is to transform algorithms
from unsigned to signed networks. {  Negative links
present distinct properties from positive
links~\cite{szell2010multirelational}. As suggested
in~\cite{leskovec2010predicting}, positive and negative links should
also be viewed as tightly related features in signed social
networks. Meanwhile links could have different  semantics in
different social media sites.  Therefore, an alternative approach is
to develop novel models based on the understandings about signed
networks.} Very recently, a framework is proposed to capture both single- and multi-view information from signed networks for node classification that significantly improves the classification performance~\cite{tang2015node}.

\subsubsection{Node Embedding}

Node embedding (or network embedding), which aims to learn
low-dimensional vector representations  for nodes, has been proven
to be useful in many tasks of social network analysis such as link
prediction\cite{liben2007link}, community
detection\cite{papadopoulos2012community}, and node
classification\cite{bhagat2011node}. The vast majority of existing
algorithms have been designed for social networks with only positive
links while the work on signed network embedding is rather limited.

Given a signed network $\mathcal{G}(\mathcal{N},{\bf A}^n, {\bf
A}^p)$,  the task of signed-network embedding is to learn a
low-dimensional vector representation ${\bf u}_i \in \mathbb{R}^d$
for each user $u_i$ where $d$ is the embedding dimension. Similar to
unsigned network embedding, a signed network embedding algorithm
needs (1) an objective function for signed network embedding; and
(2) a representation learning algorithm to optimize the objective
function. Social theories for unsigned social networks have been
widely used to design objective functions for unsigned social
network embedding. For example, social correlation theories such as
homophily and social influence suggest that two positively connected
users are likely to share similar interests, which are the
foundations of many objective functions of unsigned network
embedding~\cite{belkin2001laplacian}. Many social theories such as
balance and status theories are developed for signed social networks
and they provide fundamental understandings about signed social
networks, which could pave us a way to develop objective functions
for signed network embedding. Meanwhile recently deep learning
techniques provide powerful tools for representation learning which
have enhanced various domains such as speech recognition, natural
language processing and computer vision~\cite{yann2015}.  Therefore
a useful future direction is to harness the power of deep learning
techniques to learn low-dimensional vector representations of nodes
while preserving the fundamental understanding about signed social
networks from social theories.

\section{Link-oriented Tasks}

\begin{figure}
    \begin{center}
      \subfloat[Link Prediction]{\label{fig:pnlinkprediction}\includegraphics[scale=0.3]{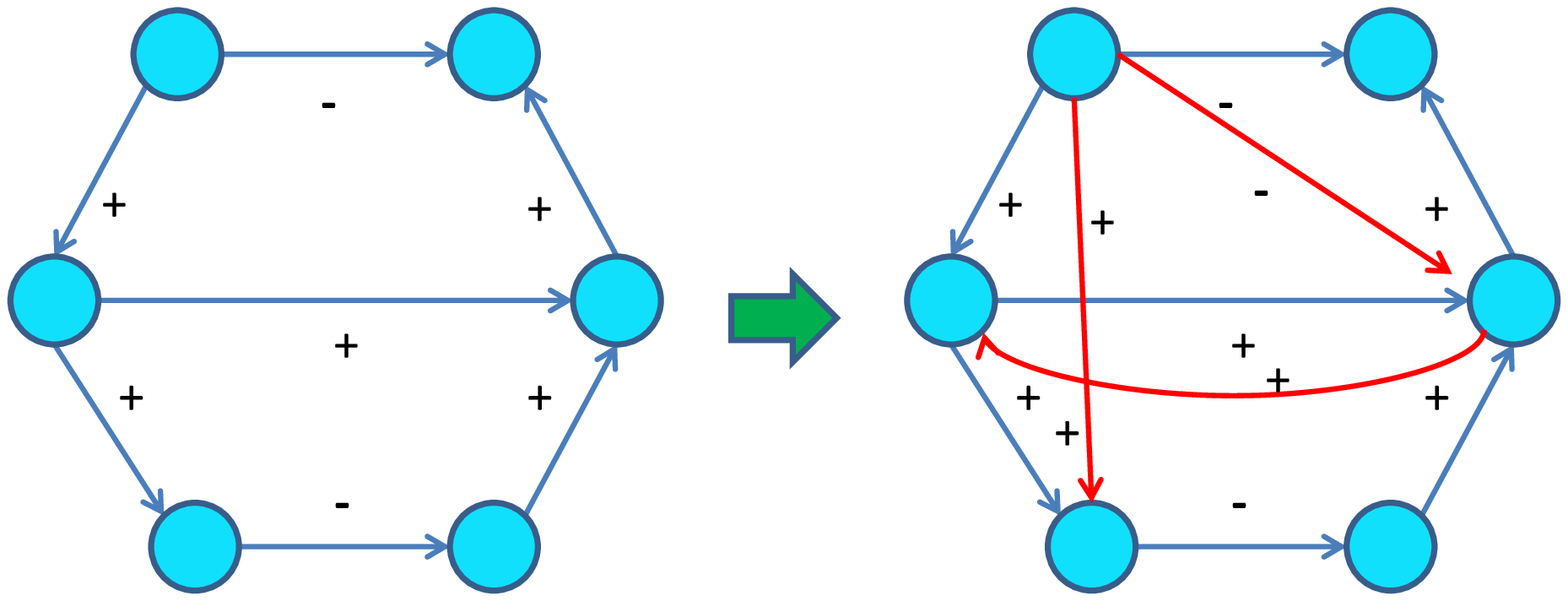}}
    \subfloat[Sign Prediction]{\label{fig:signprediction}\includegraphics[scale=0.3]{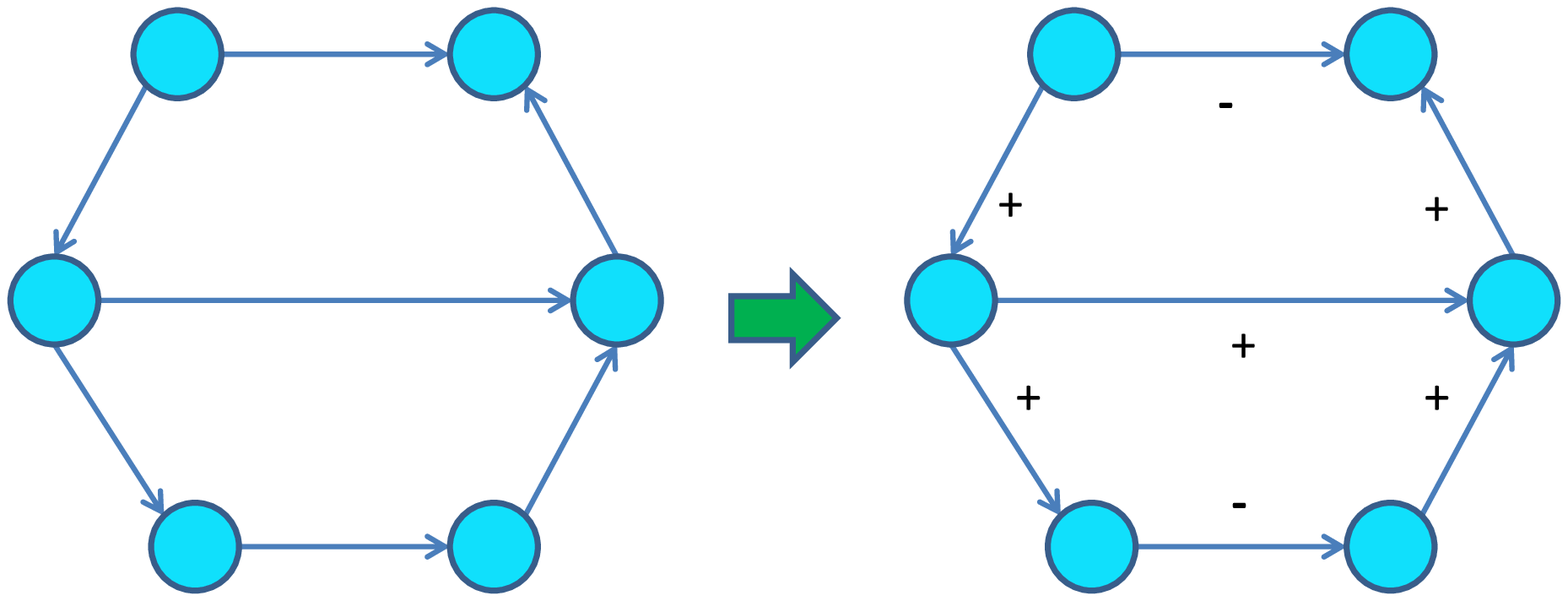}}
      \end{center}
\caption{An Illustration of the Differences of Link Prediction and Sign Prediction.}
\label{fig:pnsign}
\end{figure}

The objects of link-oriented tasks are links among nodes, which aim
to reveal fine-grained and comprehensive understandings of links.
The availability of negative links in signed networks not only
enriches the existing link-oriented tasks for unsigned networks such
as link prediction and tie strength prediction, but only encourages
novel link-oriented tasks specific to signed networks such as sign
prediction and negative link prediction. In this section, we review
two extensively investigated link-oriented tasks in signed networks
including link prediction and sign prediction. We would like  to
clarify the differences of these two tasks since they are used
interchangeably in some literature. The differences of link
prediction and sign prediction are demonstrated in
Figure~\ref{fig:pnsign}:
\begin{itemize}
\item In link prediction, signs of old links are given, while no signs are given to links in sign prediction; and
\item Link prediction predicts new positive and negative links, while sign prediction predicts signs of existing links.
\end{itemize}

\subsection{Link Prediction in Signed Networks}

Link prediction infers new positive and negative links by giving old
positive and negative
links~\cite{leskovec2010predicting,chiang2011exploiting}. Existing
link prediction algorithms can be roughly divided into two groups,
which correspond to  supervised and unsupervised methods. Supervised
methods consider the link prediction problem as a classification
problem by using the existence of links as labels, while
unsupervised methods make use of the topological properties of the
snapshot of the network. Next, we will review each group with
representative algorithms.

\subsubsection{Supervised Methods}

Supervised methods treat link prediction as a classification problem
and usually consist of two important steps. One is to prepare
labeled data and the other is to construct features for each pair of
users. The first step is trivial since the signs of links can be
naturally treated as labels. Therefore different algorithms in this
family provide different approaches to construct features.

In addition to indegree and outdegree of positive (or negative)
links, triangle-based features according to balance theory are
extracted in~\cite{leskovec2010predicting}. Since signed social
networks are usually very sparse and most users have few of indegree
or outdegree, many users could have no triangle-based features and
triangle-based features may not be
robust~\cite{chiang2011exploiting}. A link prediction algorithm can
be developed based on any quantitative social imbalance measure of a
signed network. Hence, $k$-cycle-based features are proposed
in~\cite{chiang2011exploiting}, where triangle-based features are
special cases of $k$-cycle-based features when $k = 3$. In addition
to $k$-cycle-based features, incoming local bias (or the percentage
of negative reviews it receives in all the incoming reviews) and
outgoing local bias (or the percentage of negative reviews it gives
to all of its outgoing reviews) are also reported to be helpful for
the performance improvement in link
prediction~\cite{zhang2013characterization}. In chemical and
biological sciences, the quantitative structure-activity
relationship hypothesis suggests that ``similar molecules'' show
``similar activities'', e.g., the toxicity of a molecule can be
predicted by the alignment of its atoms in the three-dimensional
space.  { This hypothesis may be applicable to social networks --
the structure and network patterns of the ego-networks are strongly
associated with the signs of their generated links. Therefore,
frequent sub-networks from the ego-networks are used as features
in~\cite{papaoikonomou2014edge}. } Besides features extracted  from
topological information, attributes of users such as gender, career
interest, hometown, movies, thinking are also used as features
in~\cite{patidar2012predicting} where it first trains a classifier
based on these features, then suggests new links and finally refines
them which either maintain or enhance the balance index according to
balance theory. Other types of features are also used for the
problem of link prediction in signed networks including user
interaction features~\cite{dubois2011predicting} and review-based
features~\cite{borzymek2010trust}. Interaction features are reported
to be more useful than node attribute features
in~\cite{dubois2011predicting}.

\subsubsection{Unsupervised methods}

Unsupervised methods are usually based on certain topological
properties of signed networks. Algorithms in this family can be
categorized into similarity-based, propagation-based, and low-rank
approximation-based methods.

\noindent {\bf Similarity-based Methods:} Similarity-based methods
predict the signs of links based on node similarity. {
Note that similarity-based methods are typically designed for
undirected signed networks.} A typical similarity-based method
consists of two steps. First,  it  defines a similarity metric to
calculate node similarities. Then,  it provides a way to predict
positive and negative links based on these node similarities.

One popular way of calculating node similarity is based on user
clustering. We discuss two representative approaches below:
\begin{itemize}
\item {  The network is partitioned into a number of clusters using the method in~\cite{doreian1996partitioning}}. Then,  the conditional
 similarity for two clusters A and B with a third cluster C is defined
 according to~\cite{javari2014cluster}:
\begin{align}
Sim_{A,B|C} = \frac{\sum_{s \in S_{A,B|C}~ m_{A,s} m_{B,s}}}{\sqrt{s \in S_{A,B|C} m_{A,s}^2}\sqrt{s \in S_{A,B|C} m_{B,s}^2}}
\end{align}
\noindent where $S_{A,B|C}$ is the set of nodes in the cluster C,
which are linked by nodes in A and B, and $m_{A,s}$ is the average
signs of links from nodes in cluster A to node $s$. Node similarity
is calculated as the similarity between clusters where these two
nodes are assigned.
\item Spectral clustering based on the Laplacian matrix for signed networks
is performed~\cite{symeonidis2013spectral}. Then,  two similarities
are defined. The first is  the similarity of nodes that are assigned
to the same cluster:
\begin{align}
simSC(i,j) = 1 - \|D(i,c_i) - D(j,c_j)\|
\end{align}
The second is  the similarity of nodes that are assigned to
different clusters:
\begin{align}
simDC(i,j) = \frac{1}{1 + D(i,c_i) + D(j,c_j)}
\end{align}
\noindent where $D(.,.)$ is a distance metric.
\end{itemize}

Another way of calculating node similarity is based on status
theory. According to status theory, the positive in-degree $|I^+|$ and
the negative out-degree $|O^-|$ of a node increase its status. In
contrast, the positive  out-degree $|O^+|$, and negative in-degree
$|I^-|$ decrease its status. According to this intuition, similarity
is defined as follows~\cite{symeonidis2013transitive}:
\begin{align}
&sim(i,j) = \frac{1}{\sigma(i)+\sigma(j) - 1} \nonumber \\
&\sigma(i) = |I_i^+| + |O_i^-| - |O_i^+| - |I_i^-|
\end{align}
With node similarity, the second step is to determine the signs of
links. Since we have pair-wise node similarities, user-oriented
collaborative filtering are used to aggregate signs from  similar
nodes to predict positive and negative
links~\cite{javari2014cluster}. {  Another approach is
based on status theory and  the sign from $i$ to $j$ is predicted as
the sign of the sum  of $sign (sim(i,k) + sim(k,j))$ over all
triplets $(i,j,k)$~\cite{symeonidis2013transitive}.}

\noindent {\bf Propagation-based Methods:} The vast majority of
propagation-based methods are proposed for trust-distrust networks,
which are a special (and important)  class of signed networks.
{ The adjacency  matrix ${\bf A}$ is very sparse and many
entries in ${\bf A}$ are zero. The basic idea of propagation-based
methods is to compute a dense matrix $\hat{\bf A}$ with the same
size of ${\bf A}$ by performing certain propagation operators on
${\bf A}$. Then the sign of a link from $u_i$ to $u_j$ is predicted
as $sign(\hat{\bf A}_{ij})$ and the likelihood is $|\hat{\bf
A}_{ij}|$. } In~\cite{guha2004propagation}, trust propagation is
treated as a repeating sequence of matrix operations, which consists
of four types of atomic trust propagations. {  These four
types are  direct propagation, trust coupling, co-citation and
transpose trust as shown in Figure~\ref{fig:trustpropogation}.} Two
strategies are studied for incorporating distrust. The first is that
of {\em one-step distrust propagation}, in which  we propagate
multiple step trust and then propagate one-step distrust. The second
is that of {\em  multiple step distrust propagation} in which trust
and distrust propagate together. One step distrust propagation often
outperforms multiple step distrust
propagation~\cite{guha2004propagation}. However, one step distrust
propagation might not converge, when the  network is dominated by
distrust links. On the other hand,  multiple step distrust
propagation may yield some unexpected
behaviors~\cite{ziegler2005propagation}. To mitigate these two
problems, Ziegler and Lausen~\citeyear{ziegler2005propagation}
propose to integrate distrust into the process of the Appleseed
trust metric computation instead of superimposing distrust
afterwards. Methods in~\cite{guha2004propagation}
and~\cite{ziegler2005propagation} are based on the matrix
representation. There are methods in this family investigating other
representations such as subjective logic~\cite{knapskog1998metric},
intuitionistic fuzzy relations~\cite{de2006many} and
bilattice~\cite{victor2006towards}, which can naturally perform both
trust and distrust propagation by defining corresponding operators.
{
\begin{figure}
    \begin{center}
   \includegraphics[scale=0.5]{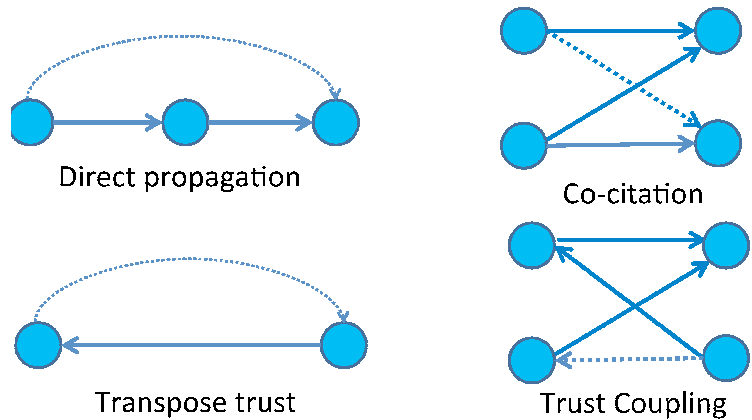}
    \end{center}
\caption{Four types of atomic trust propagations.}
\label{fig:trustpropogation}
\end{figure}
}

\noindent {\bf Low-rank approximation methods:} The notion of
balance is generalized by Davis in~\citeyear{davis1967clustering} to
weak balance, which allows triads with all negative links. Low-rank
approximation methods are based on weak structural balance as
suggested in~\cite{hsieh2012low} that weakly balanced networks have
a low-rank structure and weak structural balance in signed networks
naturally suggests low-rank models for signed networks.
{ Low-rank approximation methods compute the dense matrix $\hat{\bf A}$ via the
low-rank approximation of ${\bf A}$ instead of propagation operators for propagation-based methods.
With $\hat{\bf A}$, the sign and the likelihood of a link from $u_i$ to $u_j$ are predicted
as $sign(\hat{\bf A}_{ij})$ and $|\hat{\bf A}_{ij}|$, respectively. } In~\cite{hsieh2012low}, {  the link prediction problem in signed
networks is mathematically modeled as a low-rank matrix
factorization problem as follows:
\begin{align}
\min_{{\bf W}, {\bf H} } \sum_{i,j} {( {\bf A}_{ij} - ({\bf W}^T {\bf H})_{ij} )^2} + \lambda (\| {\bf W} \|_F^2 + \| {\bf H}\|_F^2)
\end{align}
\noindent where ${\bf W}^T {\bf H}$ is the low-rank matrix to
approximate ${\bf A}$. The square function is chosen as the loss
function in $( {\bf A}_{ij} - ({\bf W}^T {\bf H})_{ij} )^2$. }
Pair-wise empirical error, similar to the hinge loss convex
surrogate for 0/1 loss in classification, is used
in~\cite{agrawal2013link}. They use of  this particular variation
since it elegantly captures the correlations amongst the users and
thereby makes the technique more robust to fluctuations in
individual behaviors. In~\cite{cen2013sign}, a low-rank tensor model
is proposed for link prediction in dynamic signed networks.

\subsection{Sign Prediction}

Most social media services provide unsigned social networks such as
the friendship network in Facebook and the following network in
Twitter, while only few services provide signed social networks. The
task of sign prediction is to infer the signs of existing links in
the given unsigned network. It is difficult, if not impossible, to
predict signs of existing links by only utilizing the given unsigned
network~\cite{yang2012friend}. Therefore,
most of the existing sign predictors use additional sources of
information. The most widely used sources are user interaction
information and cross-media information.

\subsubsection{Sign Prediction with Interaction Data}

In reality, we are likely to adopt the opinions from our friends
while fighting the opinions of our foes. As a consequence, decisions
of users with positive links are more likely to agree, whereas for
users with negative connections, the chance of disagreement would be
considerably higher. In social media, users can perform positive or
negative interactions with other users. Positive interactions show
agreement and support, while negative interactions show disagreement
and antagonism. There are strong correlations between positive (or
negative) links and positive (or negative)
interactions~\cite{yang2012friend}.  Tang et al. suggest a
straightforward algorithm for sign prediction based on the
correlation between interactions and links. The first step is to
initialize  signs of links based on interactions. Positive signs are
used for positive interactions, whereas  negative signs are used for
negative interactions.  Next, the signs  of links are refined
according to status theory or balance theory~\cite{Tang-etal15a}.
More sophisticated algorithms incorporate link and interaction
information into coherent frameworks. In~\cite{yang2012friend}, a
framework with a set of latent factor models is proposed to infer
signs for unsigned links, which capture user  interaction behavior,
social relations as well as their interplay. It also models the
principles of balance and status theories for signed networks. A
one-dimensional latent factor $\beta_i$ is introduced for $u_i$ and
then we model the sign between $u_i$ and $u_j$ as $s_{ij} = \beta_i
\beta_j$, which can capture balance theory. The vector parameter
$\eta$ is introduced for users to capture their partial ordering, {
and then the status $\ell_i$ of $u_i$ is modeled as $\ell_i = \eta
\gamma_i$ where $\gamma_i$ is the latent factor vector of $u_i$.}
Status theory characterizes the sign from $u_i$ to $u_j$ as their
relative status difference $\ell_{ij} = \ell_i - \ell_j$. Yu and Xie
find significant correlations and mutual influence between user
interactions and signs of links. They propose a mutual latent random
graph framework to flexibly model evidence from user interactions
and signs. This approach is used to perform user interaction
prediction and sign prediction
simultaneously~\cite{yu2014modeling,yu2014learning}.

\subsubsection{Sign Prediction with Cross-Media Data}

In the task of link prediction in signed networks  Leskovec et al.
find that the learned link predictors have very good generalization
power  across social media sites. This observation  suggests that
general guiding  principles might exist for sign inference across
different networks, even when links have different semantic
interpretations in different networks~\cite{leskovec2010predicting}.
Another useful source for sign prediction is cross-media
information. The goal is to predict signs of a target network with a
source signed network. The basic approach is to learn knowledge or
patterns from the source signed network, and use it  to predict link
signs in the target network. The vast majority of  algorithms in
this family use transfer learning to achieve this goal. One
representative way is to construct generalizable features that can
transfer patterns from the source network to the target  network for
sign prediction. Since some social theories such as status and
balance theories are applicable for all signed networks, it is
possible to extract generalizable features suggested by social
theories, such as balance and status theory.
{  In~\cite{tang2012inferring}, a factor-graph model is learned with}
features from the source network to infer signs of the target
network. Although links in different signed networks may have
different semantics,  a certain degree of similarity always exists
across domains, e.g., similar degree distributions and diameters.
With this intuition, an alternative way is to project the source and
target networks into the same latent space. Latent topological
features are constructed to capture the common patterns between the
source and target networks. This is  obtained through the following
optimization problem~\cite{ye2013predicting}: {
\begin{align}
\min_{{\bf U}_s, \Sigma, {\bf V}_s, {\bf U}_t, {\bf V}_t} \| {\bf A}_s - {\bf U}_s \Sigma {\bf V}_s^\top\|_F^2 + \|{\bf A}_t - {\bf U}_t \Sigma {\bf V}_t^\top\|_F^2 + \alpha \|\Sigma\|_F^2
\end{align}
\noindent where ${\bf A}_s$ and ${\bf A}_t$ denote the adjacency
matrices for the source  and  target network, respectively. ${\bf
U}_s$, ${\bf V}_s$, ${\bf U}_t$ and ${\bf V}_t$ are four latent topological feature matrices~\cite{ye2013predicting}.} $\Sigma$ is the common
latent space for both networks, which ensures that the extracted
topological features of both graphs are expressed in the same space.
With the latent topological features, a transfer learning with
instance weighting algorithm is proposed to predict signs of the
target unsigned network ${\bf A}_t$ by learning knowledge from the
source signed network ${\bf A}_s$.

\subsection{Promising Directions for Link-oriented Tasks}
For many social media sites, negative links are usually unavailable,
which might limit the applications of mining signed networks.
Therefore, it is helpful to  predict negative links.  Furthermore,
for most signed social networks in social media, only binary
relations are available and strengths of the relations are not
available. In other words, we would like to perform {\em  tie
strength prediction}. In this subsection, we discuss these two
link-oriented tasks.

\subsubsection{Negative Link Prediction}

\begin{figure}
    \begin{center}
      \includegraphics[scale=0.3]{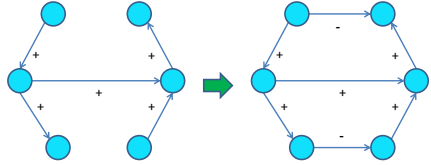}
      \end{center}
\caption{An Illustration of The Problem of Negative Link Prediction.}
\vspace*{-0.15in}
\label{fig:nelp}
\end{figure}

It is evident from recent work that negative links have significant
added value over positive links in various analytical tasks such as
positive link
prediction~\cite{guha2004propagation,leskovec2010predicting}, and
recommender systems~\cite{victor2009trust,ma2009learning}. However,
it is generally not very desirable for online social networks to
explicitly collect negative
links~\cite{hardin2004distrust,kunegis2013added}. As a consequence,
the vast majority of social media sites such as Twitter and Facebook
do not enable users to explicitly specify negative links. Therefore,
it is natural to question whether one can predict negative links
automatically from the available data in social networks. While this
problem is very challenging~\cite{chiang2013prediction}, the results
of such an approach  have the potential to improve the quality of
the results of a vast array of applications. The negative link
prediction problem is illustrated in Figure~\ref{fig:nelp}. { The
negative link prediction problem is different from both the link prediction and sign prediction problems as follows:}

\begin{itemize}
\item Link prediction in signed networks
predicts positive and negative links from existing positive and negative links.
On the other hand, negative link prediction does not assume the existence of negative links.
\item Sign prediction predicts signs of {\em already existing} links. While the negative link prediction problem needs to identify the pairs of nodes between which negative links {\em are predicted to} exist.
\end{itemize}

A recent work in~\cite{Tang-etal15a} found that negative links can
be predicted with user interaction data by using the correlation
between negative interactions and negative links. Furthermore, the
proposed negative link predictor in~\cite{Tang-etal15a} has very
good generalization across social media sites, which suggests that
cross-media data might be also helpful in the problem. It is
possible to build signed networks via sentiment analysis of
texts~\cite{hassan2012extracting,wang2014recognizing}, which
suggests that user-generated content has significant potential in
predicting negative links in social media.

\subsubsection{Tie-Strength Prediction}
The cost of forming links in social media is very low, as a result
of which many weak ties are formed~\cite{xiang2010modeling}.
 The authors of~\cite{huberman2008social} show that
users can have many followees and followers in Twitter with whom
they are only weakly associated  in the physical world.  Users with
strong ties tend to be  more similar  than those with weak ties.
Since homophily is a useful property from the perspective of mining
tasks, such as recommendation and friend management,  it suggests
that tie-strength prediction can also be very useful. For unsigned
networks in social media, such as friendship in Facebook and
Twitter, we often choose a binary adjacency matrix representation
where $1$ denotes a positive link from $u_i$ to $u_j$ and 0
otherwise. The tie-strength prediction task in unsigned networks is
to infer a strength in $[0,1]$ for a given positive link. The
original binary matrix representation with values in $\{0,1\}$ is
converted into a continuous valued matrix representation with values
in $[0,1]$ by tie-strength prediction in unsigned networks.

If we choose one adjacency matrix ${\bf A}$ to represent a signed
network with $\{-1,0,1\}$ to denote negative, missing and positive
links, a tie strength predictor infers strength values in [-1,0] for
negative links and [0,1] for positive links. If we choose two
adjacency matrices ${\bf A}^p$ and ${\bf A}^n$ in $\{0,1\}$ to
represent positive and negative links separately, a tie strength
predictor infers strength values in [0,1] for positive and negative
links.

Previous studies  in positive tie-strength prediction problem
suggest that  pairwise user similarity is reflected in strong ties.
Therefore,  the strengths of positive ties are modeled as the hidden
impacts of node similarities. Furthermore,  the strengths of
positive ties are modeled as the hidden causes of user interactions
since they affects the frequency and nature of user
interactions~\cite{xiang2010modeling}. A preliminary work
in~\cite{Tang-etal14d} finds that it is more likely for two users to
have  negative links if they have more negative interactions.
Analogously, this suggests the following directions for tie-strength
prediction: (a) What is the relation between negative tie strength
and node-node similarities and how negative tie strength impacts
user interactions; and (b) how negative and positive tie strength
affect one another.

\section{Application-oriented Tasks}

Just as  unsigned networks are used frequently in the context of
various applications  such as data
classification~\cite{zhu2007combining}, data
clustering~\cite{long2006spectral}, information
propagation~\cite{kempe2003maximizing} and
recommendation~\cite{tang2013social}, signed networks can be
leveraged as well. Application-oriented tasks augment traditional
algorithms with signed networks. For example, in addition to rating
information, recommender systems with signed networks can also make
use of signed networks. In this section, we review the
recommendation and information diffusion applications and discuss
promising research directions.

\subsection{Recommendation with Signed Networks}

Assume that ${\bf R}$ is the user-item ratings matrix where ${\bf
R}_{ij}$ is the rating from the $i$-th user to the $j$-th item. In a
typical recommender system, most of the entries are missing.
Traditional recommender systems aim to predict these missing values
by using observed values in ${\bf R}$. In the physical world, we
always seek recommendations from our friends, which suggests that
social information may be useful to improve recommendation
performance. Many recommender systems are proposed to incorporate
ones' friends for recommendation and gain performance improvement. A
comprehensive review about social recommendation can be found
in~\cite{tang2013social,tang2014recommendation}. Scholars have noted
that negative links may be more noticeable and credible than the
positive links with a similar magnitude~\cite{cho2006mechanism}.
Negative links may be as important as positive links for
recommendation. In recent years, systems based on collaborative
filtering (CF) are proposed to incorporate both positive and
negative links for recommendation. Typically, a CF-based recommender
system with signed networks contains two components corresponding to
the basic CF model and the model extracted from the signed network.
The  basic CF models are categorized  into memory-based and
model-based systems.

\subsubsection{Memory-based methods}

Memory-based recommender systems with signed networks choose
memory-based collaborative filtering,  and especially user-oriented
models~\cite{victor2009trust,victor2013enhancing,chen2013effective,nalluriutility}.
{  A typical user-oriented model first calculates pair-wise
user similarity based on some similarity metrics such as Pearson's
correlation coefficient or cosine similarity}. Then,  a missing
rating of user $i$ for item $j$  is predicted by aggregating ratings
from the similar peers of user $i$ as follows:
\begin{align}
\hat{{\bf R}}_{ij} = \hat{r}_i + \frac{\sum_{v \in N_i} W_{iv} ({\bf R}_{vj} - \hat{r}_v)}{\sum_{v \in N_i} W_{iv} }
\end{align}
\noindent where $N_i$ is the set of similar users of $u_i$,
$\hat{r}_i$ is the average rating from $u_i$ and $W_{iv}$ is the
connection strength between $u_i$ and $u_v$. There are several
strategies for incorporating  negative links into the above
user-oriented model as:
\begin{itemize}
\item One is to use negative links to avoid recommendations from these ``unwanted'' users as~\cite{victor2009trust}:
\begin{align}
\hat{{\bf R}}_{ij} = \hat{r}_i + \frac{\sum_{v \in N_i\backslash D_i} W_{iv} ({\bf R}_{vj} - \hat{r}_v)}{\sum_{v \in N^+} W_{iv} }
\end{align}
\noindent $D_i$ is the set of users to whom $u_i$ has negative links.
\item Another way is to consider negative links as negative weights, i.e., considering negative links as dissimilarity measurements, as~\cite{victor2013enhancing}:
\begin{align}
\hat{{\bf R}}_{ij} = \hat{r}_i + \frac{\sum_{v \in N_i} W_{iv} ({\bf R}_{vj} - \hat{r}_v)}{\sum_{v \in N_i} W_{iv} } - \frac{\sum_{v \in D_i} d_{iv} ({\bf R}_{vj} - \hat{r}_v)}{\sum_{v \in D_i} d_{iv} }
\end{align}
\noindent where $d_{iv}$ is the dissimilarity between $u_i$ and $u_v$.
\item In reality, positive and negative links in signed networks are very sparse therefore Nalluri proposes a recommender system, which first propagates positive and negative values in signed networks and then reduces the influence from negative values as~\cite{nalluriutility}:
\begin{align}
\hat{{\bf R}}_{ij} = \hat{r}_i + \frac{\sum_{v \in N_i} (W_{iv} - d_{iv}) ({\bf R}_{vj} - \hat{r}_v)}{\sum_{v \in N_i} (W_{iv} - d_{iv}) }
\end{align}
\end{itemize}

\subsubsection{Model-based Methods}

Model-based recommender systems with negative links use model-based
collaborative filtering.  Matrix factorization methods are
particularly popular~\cite{ma2009learning,forsati2014matrix}. Assume
that ${\bf U}_i$ is the $k$-dimensional preference latent factor of
$u_i$ and ${\bf V}_j$ is the $k$-dimensional characteristic latent
factor of item $j$. A typical matrix factorization-based
collaborative filtering method models the rating from $u_i$ to the
$j$-th item ${\bf R}_{ij}$ as the interaction between their latent
factors, i.e., ${\bf R}_{ij} = {\bf U}_i^\top {\bf V}_j$ where ${\bf
U}_i$ and ${\bf V}_j$ can be obtained by solving the following
optimization problem:
\begin{align}
\min_{{\bf U},{\bf V}} \sum_{i=1}^n \sum_{j=1}^m {\bf W}_{ij}({\bf R}_{ij} - {\bf U}_i {\bf V}_j^\top)^2 + \alpha (\|{\bf U}\|_F^2 + \|{\bf V}\|_F^2),
\label{eq:recmf}
\end{align}
\noindent where {  ${\bf U} = [{\bf U}_1^\top, {\bf U}_2^\top,
\ldots, {\bf U}_N^\top]^\top\in \mathbb{R}^{n\times K}$ and ${\bf V}
= [{\bf V}_1^\top, {\bf V}_2^\top, \ldots, {\bf V}_M^\top]^\top \in
\mathbb{R}^{m\times{K}}$ where $N$ and $M$ are the numbers of users
and items in a recommender system}. The term $\|{\bf U}\|_F^2 +
\|{\bf V}\|_F^2$ is introduced to avoid over-fitting, controlled by
the parameter $\alpha$. ${\bf W} \in \mathbb{R}^{n\times m}$ is a
weight matrix where ${\bf W}_{ij}$ is the weight for the rating for
$u_i$ to $v_j$. A common way to set ${\bf W}$ is ${\bf W}_{ij} = 1$
if we observe a rating from $u_i$ to the $j$-th item, and ${\bf
W}_{ij} = 0$ otherwise. {The optimization problem in
Eq.~(\ref{eq:recmf}) is convex for ${\bf U}$ and ${\bf V}$,
respectively.  Therefore it is typically solved by gradient decent
methods or alternating least squares.}  If $u_i$ positively link to
$u_j$, $u_i$ and $u_j$ are likely to share similar preferences.
Therefore, to capture positive links, Ma et
al.~\cite{ma2011recommender} added a term to minimize the distance
of the preference vectors of two users with a positive link based on
Eq.~(\ref{eq:recmf}) as follows: {
\begin{align}
\min_{{\bf U},{\bf V}} \sum_{i=1}^n \sum_{j=1}^m {\bf W}_{ij}({\bf R}_{ij} - {\bf U}_i {\bf V}_j^\top)^2 + \alpha (\|{\bf U}\|_F^2 + \|{\bf V}\|_F^2) + \beta \sum_i \sum_{j \in N_i} S_{ij}^p \|{\bf U}_i - {\bf U}_j\|_2^2
\label{eq:nt}
\end{align}}
\noindent where  $S_{ij}^p$ is the strength of the positive link
from $u_i$ to $u_j$, and $\beta$ controls the contribution from
positive links.

If $u_i$ has a negative link to $u_j$,  it is likely that  $u_i$
thinks that $u_j$ has totally different tastes. With this intuition,
for a negative link from $u_i$ to $u_j$, Ma et
al.~\cite{ma2009learning} introduce a term to maximize the distance
of their latent factors based on the matrix factorization model as
follows:
\begin{align}
\min_{{\bf U},{\bf V}} \sum_{i=1}^n \sum_{j=1}^m {\bf W}_{ij}({\bf R}_{ij} - {\bf U}_i {\bf V}_j^\top)^2 + \alpha (\|{\bf U}\|_F^2 + \|{\bf V}\|_F^2)  - \beta \sum_i \sum_{j \in D_i} S_{ij}^n \|{\bf U}_i - {\bf U}_j\|_2^2
\label{eq:ndis}
\end{align}
\noindent where $S_{ij}^n$ is the strength of the negative link for
$u_i$ to $u_j$. The underlying assumption of Eq.~(\ref{eq:ndis}) is
to consider negative links as dissimilarity measurements. { Gradient descent is
performed in~\cite{ma2009learning} to obtain a local minimum of the objective function given by Eq.~(\ref{eq:ndis})}. However, recent research suggests that negative links may not denote
dissimilarity and users with negative links tend to be more similar
than randomly selected pairs~\cite{Tang-etal14b}. It also observes
that users with positive links are likely to be more similar than
pairs of users with negative links, which is very consistent with
the extension of the notion of structural balance
in~\cite{cygan2012sitting} -- a structure in signed network should
ensure that users are able to have their ``friends`` closer than
their ``enemies'', i.e., users should sit closer to their
``friends'' (or users with positive links) than their ``enemies''
(or users with negative links). With this intuition, for $\langle
i,j,k\rangle$ where $u_i$ has a positive link to $u_j$ while has a
negative link to $u_k$, the latent factor of $u_i$ should be more
similar to the latent factor of $u_j$ than that of $u_k$ to capture
negative links. In particular, for each triple as $\langle
i,j,k\rangle$, a regularization term is added as follows:
\begin{align}
\ell(d({\bf U}_i, {\bf U}_j), d({\bf U}_i, {\bf U}_k))
\label{eq:fe}
\end{align}
\noindent where $d$ is a similarity metric and $\ell$ is a penalty
function that assesses the violation of latent factors of users with
positive and negative links~\cite{forsati2014matrix}. Possible
choices of $\ell(z)$ are the hinge loss function $\ell(z) =
\max(0,1-z)$ and the logistic loss function $\ell(z) =
\log(1+e^{-z})$. {  In~\cite{forsati2014matrix}, stochastic gradient
descent (SGD) method is employed to optimize Eq.~(\ref{eq:fe}).} For
a signed network with $N$ users, there could be $N^3$ triples that
indicates we need to introduce $N^3$ possible regularization terms
as Eq.~(\ref{eq:fe}) to capture the signed network for
recommendations~\cite{forsati2014matrix}.  Therefore, the
computational cost of the system is very high.
In~\cite{tang2016recommendations}, a system with only $N$ extra
regularization terms is proposed that is much more efficient.  A
sophisticated recommender system is proposed
in~\cite{yang2012friend}. This system has several advantages -- (1)
it can perform recommendation and sign prediction simultaneously;
and (2) it is the first framework to model balance theory and status
theory explicitly for recommendation with signed networks.

\subsection{Information Diffusion}

Information diffusion can enable various online applications such as
effective viral marketing and has attracted increasing
attention~\cite{kempe2003maximizing,chen2009efficient}. There are
many information diffusion models for unsigned social networks
including the classic voter model~\cite{clifford1973model},
susceptible-infected-recovered (SIR) epidemic
model~\cite{may2001infection}, independent cascade (IC)
model~\cite{goldenberg2001talk,goldenberg2001using}, and the
threshold
model~\cite{granovetter1978threshold,schelling2006micromotives}. One
can apply these models of unsigned networks to signed networks by
ignoring negative links. However, ignoring negative links might
result in over-estimation of the impact of positive
links~\cite{li2013influence}.  {  Therefore studying information
diffusion and maximization in signed networks can not only help us
understand the impact of user interactions on information diversity
but also can push the boundaries of researches about dynamical
process in complex networks. In addition, empirical results on
real-world signed networks demonstrate that incorporating link signs
into information diffusion models usually gains
influence~\cite{li2013influence,li2014polarity,shafaei2014community}.
For example, the voter model with negative links generates at
maximum of $38\%$ and $21\%$ more influence in the Epinions dataset
compared to the model with only positive
links~\cite{li2013influence}. In the rest of this section, we will
review representative diffusion models for signed networks}

\subsubsection{Voter Model for Signed Networks}

A typical scenario of the application of the voter model is when
users' opinions switch forth and back according to their
interactions with other users in networks. The authors
of~\cite{li2013influence,li2014voter} investigate how two opposite
opinions diffuse in signed networks based on the voter model
proposed in~\cite{clifford1973model}. It is more likely for users to
adopt and trust opinions from their friends, while users are likely
to adopt the opposite opinions of their foes. This intuition
corresponds to the principles of ``enemies' enemies are my friends''
and ``my enemies' friends are my enemies''. Hence, each node $u_i$
selects one user $u_j$ from his/her outgoing social networks
randomly and performs two possible actions -- (1) if $u_i$ has a
positive link to the selected user $u_j$, $u_i$ adopts $u_j$'s
opinion; and (2) if $u_i$ has a negative link to $u_j$, $u_i$
chooses the opinion opposite to $u_j$'s.

\subsubsection{Susceptible-infected-recovered (SIR) Epidemic Model for Signed Networks}

Using epidemiology to study information spread has become
increasingly popular in recent years~\cite{may2001infection} because
the information spread mechanisms are qualitatively similar to those
of the biological disease spread~\cite{volz2007susceptible}. The
standard susceptible-infected-recovered (SIR) model assigns one of
three states (susceptible, infected, or recovered) to each user.
Based on SIR, the authors of~\cite{li2013opinion,fan2012analysis}
define five states for signed networks -- (1) $S_0$: susceptible
with neutral opinions; (2) $I_-$: infected with negative opinions;
(3) $I_+$: infected with positive opinions; (4) $R_-$: recovered
with negative opinions; and (5) $R_+$: recovered with positive
opinions. Users with $S_0$ can be infected by users with $I_-$ or
$I_+$; and users with $R_+$ or $R_-$ do not spread their opinions
any more. With the same intuition in~\cite{li2013influence},  users
are likely to adopt and trust opinions from their friends, while
adopting the opposite opinions of their foes. At each step, users
with state $I_+$ (or $I_-$) pick up one user from their social
networks to interact with, and they can perform four possible
actions depending on probabilities and the sign of links as shown in
Table~\ref{tab:SIR}.

\begin{table}
\centering
\caption{SIR for Signed Networks.}
\label{tab:SIR}
\begin{tabular}{|c|c|c|}
\hline
                            Actions & Probabilities & Relationship \\ \hline
$I_+ + S_0 \rightarrow I_+ + I_+$ & $\lambda_s$ & $+ 1$ \\
$I_+ + S_0 \rightarrow I_+ + I_-$ & $\lambda_o$ & $- 1$ \\
$I_- + S_0 \rightarrow I_- + I_-$ & $\lambda_s$ & $+ 1$ \\
$I_- + S_0 \rightarrow I_- + I_+$ & $\lambda_o$ & $- 1$ \\\hline
\end{tabular}
\end{table}

\subsubsection{Independent Cascade Model for Signed Networks}

Nodes in the network are assigned  one of two states, active or
inactive, by independent cascade model~\cite{goldenberg2001talk}. At
the $t$-th step, every active node $u_i$ has one single opportunity
to activate inactive users $u_j$ in his/her network with an
independently successful probability $p_{ij}$. $u_j$ becomes active
in the $t+1$-th step if $u_i$ succeeds. After this opportunity,
$u_i$ cannot take actions on $u_j$ any more in subsequent steps. The
authors of~\cite{li2014polarity} propose a Polarity-related
Independent Cascade (ICP) model for signed networks. Each node in {
the ICP model is assigned to one of three states }-- (1) negative:
adopting but opposing the spreading opinion, (2) positive: adopting
and supporting the opinion, and (3) inactive: not adopting the
opinion. There are two major differences between ICP model and the
standard IC model. First, each user can be only activated once in
each step for ICP. Second, if $u_i$ activates $u_j$, the state $S_j$
of $u_j$ depends on $u_i$'s state $S_i$ and the sign of their link
as $S_j = S_i \times s_{ij}$

\subsubsection{Threshold Model for Signed Networks}

The node $u_i$ becomes active in the threshold model if and only if
his/her active neighbors are more than a threshold $\theta_i$ as -
$\sum_{\text{ $u_j$ active neighbor of $u_i$}} b_{ij} > \theta_i$
where $b_{ij}$ is a weight between $u_i$ and $u_j$. The authors
of~\cite{shafaei2014community} introduce an information diffusion
model based on the threshold model for signed networks where each
node maintains a payoff matrix. { If the payoff matrices for all nodes are the same, the
proposed model boils down to the standard threshold model.} We assume
that there are two behaviors ``B'' and ``A''; all nodes start with
``B'' and then some randomly selected nodes change to ``A''. In each
iteration, every node observes his/her social network, calculates
the payoff matrix and then adopts the behavior maximizing the
benefits to him/her. Note that the payoff matrix is calculated only
for these nodes with behavior ``B''. If many friends have the
same behavior, doing the behavior changes can increase the payoff
gain, which also increases if few  foes have the behavior.

\subsection{Promising Directions for Application-oriented Tasks}

Unsigned networks are exploited to help various real-world
applications such as data classification~\cite{sindhwani2005beyond},
data clustering~\cite{long2006spectral}, active
learning~\cite{bilgic2010active}, information
propagation~\cite{kempe2003maximizing},
recommendation~\cite{tang2013social}, sentiment
analysis~\cite{speriosu2011twitter} and feature
selection~\cite{tang2012feature}.  Therefore, there are many
opportunities in the signed network setting.  In this subsection, we
focus our discussions on two application-oriented tasks, which are
data classification and clustering. We focus on these tasks because
{these problems are very general  and have applicability to}
many problems such as  sentiment
analysis~\cite{tan2011user,hu2013exploiting}. Furthermore, we can
follow similar ways for data classification and clustering problems
to define other application-oriented tasks such as active learning
and feature selection.

\begin{figure}
    \begin{center}
      \subfloat[Data with Signed Networks] {\label{fig:socialmediadata}\includegraphics[scale=0.28]{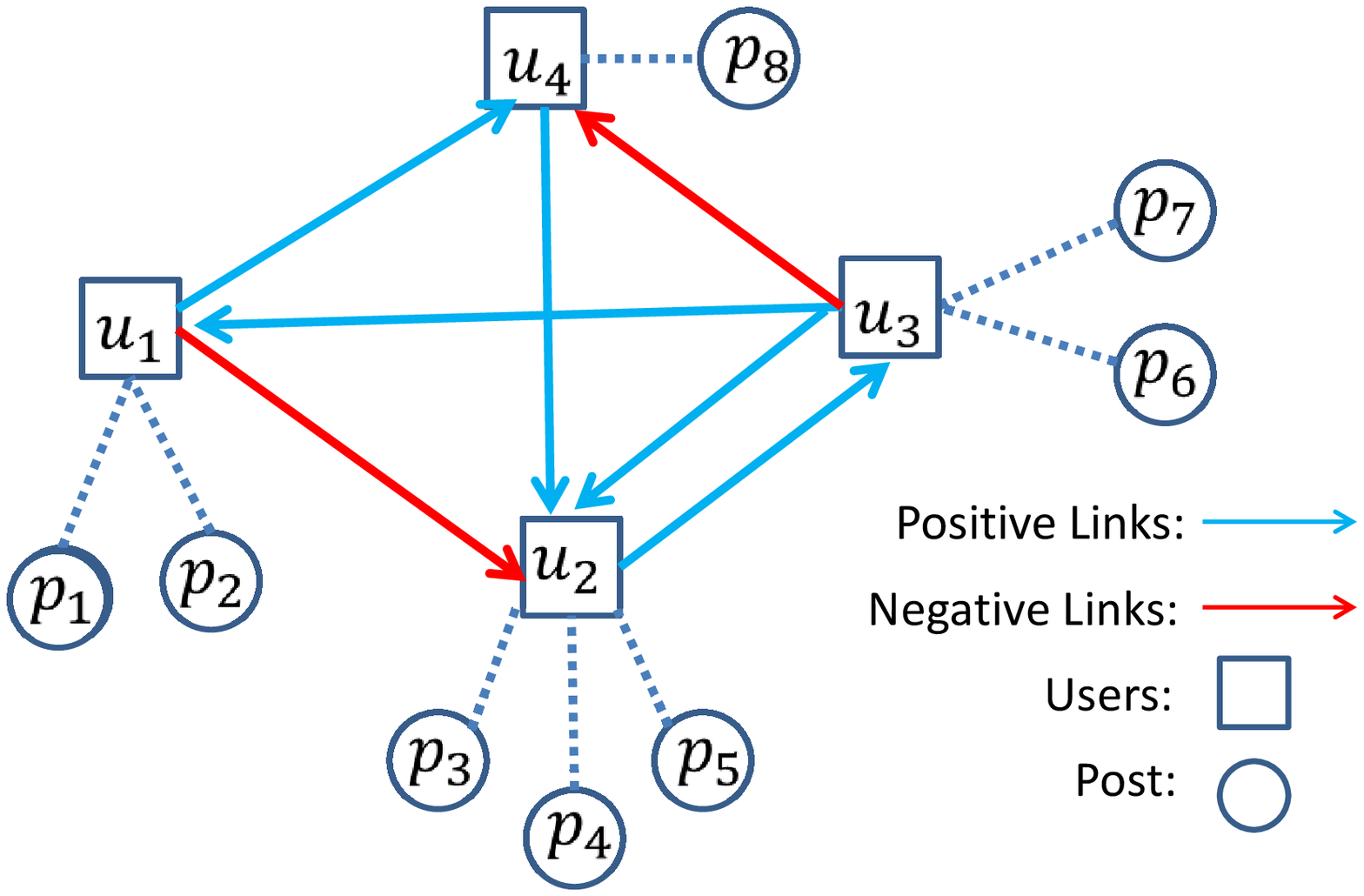}}
      \subfloat[Attribute-Value Data]{\label{fig:iid}\includegraphics[scale=0.28]{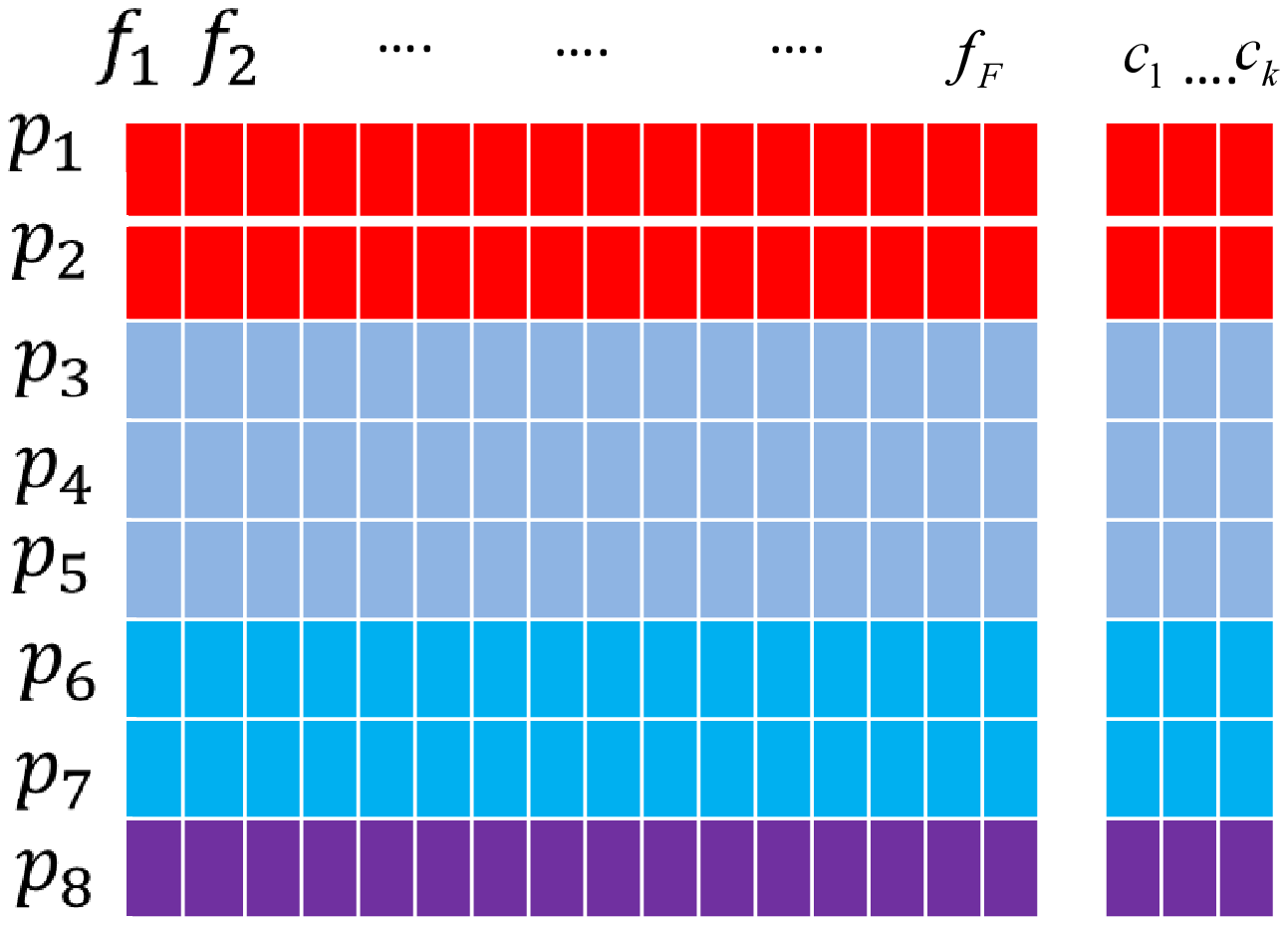}}
    \subfloat[Representation with Signed Networks]{\label{fig:dependent}\includegraphics[scale=0.28]{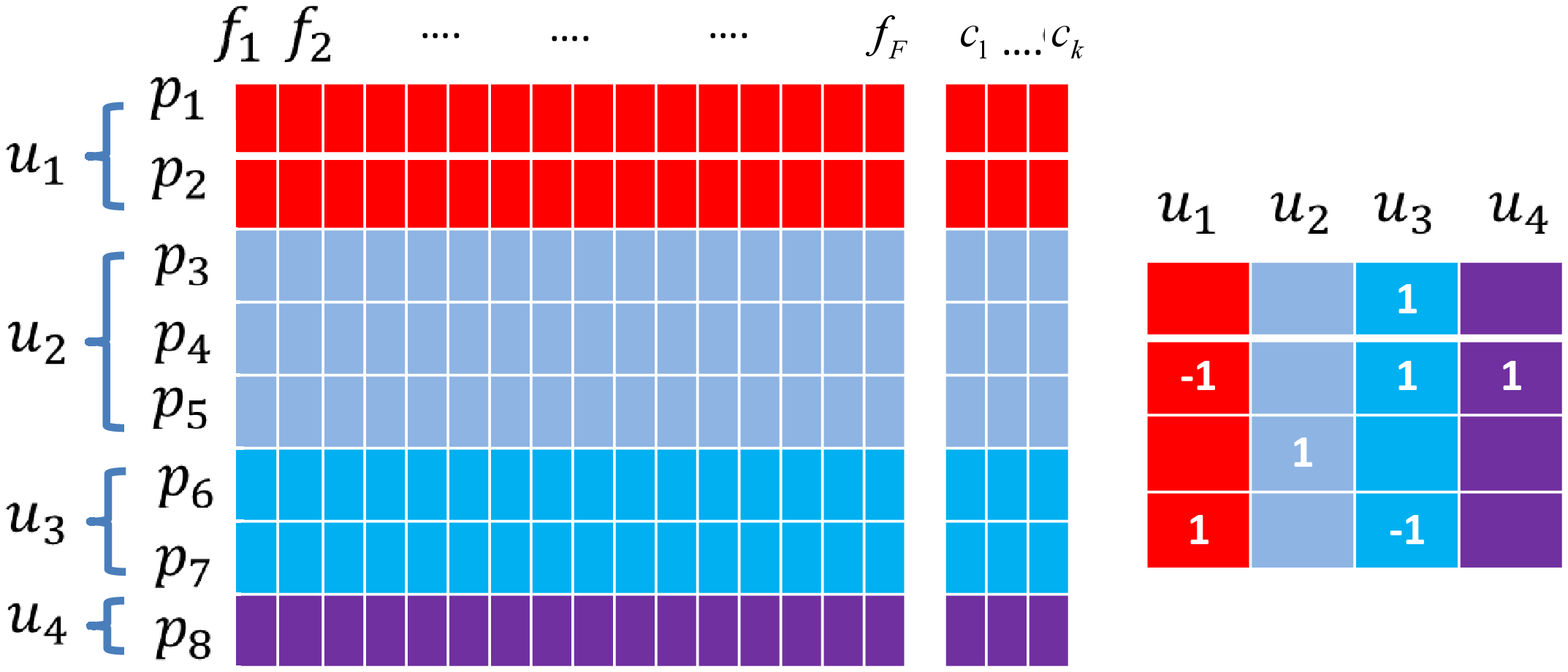}}
    \end{center}
\caption{A Simple Example of Data in the Problem of Data Classification with Signed Networks}
\label{fig:socialmediadata-represenations}
\end{figure}

\subsubsection{Data Classification with Signed Networks}

Figure~\ref{fig:socialmediadata-represenations} demonstrates a
simple example for data classification with signed networks. The
signed network in Figure~\ref{fig:socialmediadata} has four users
($u_1, \ldots, u_4$) and each user has some posts (e.g., $u_1$ has
two posts $p_1$ and $p_2$). We use posts in a loose way to cover
various types of user-generated content such as posts, tweets, or
images. In data classification with signed networks, there is
additional link information such as user-post and user-user links as
shown in Figure~\ref{fig:dependent}. Let $\mathcal{F} =
\{f_1,f_2,\ldots,f_F\}$ be a set of $F$ features and $\mathcal{P} =
\{p_1,p_2,\ldots,p_M\}$ be the set of $M$ posts. ${\bf P} \in
\mathbb{R}^{N\times M}$ denotes the user-post authorship matrix
where ${\bf P}_{ij} = 1$ if $u_i$ creates $p_j$, and $0$ otherwise;
${\bf X} \in \mathbb{R}^{M \times F}$ {  denotes the attribute-value
representation of $\mathcal{P}$ and ${\bf Y} \in \mathbb{R}^{M
\times c}$ is the label indicator matrix where ${\bf Y}_{ij} = 1$ if
$p_i$ is labeled as the $j$-th class and $0$ otherwise.} {\it  The
problem of data classification with signed networks  is  that of
training classifiers to predict class labels for unseen posts by
utilizing data instances $\{ {\bf X}, {\bf Y}\}$ and their
contextual information from signed networks $\{{\bf P}, {\bf A}\}$}.

Research on data classification with unsigned networks found that
class labels of posts from the same user are likely to be
consistent, and that users with links are likely to generate posts
with similar class labels~\cite{tang2012feature}. There are two
popular ways of exploiting contextual information from unsigned
networks for data classification based on these two findings. One
way is  to convert contextual information into correlation links
between posts.  This boils down to the problem of combining content
and correlation links for data classification~\cite{qi2009web}. The
other way is that we first extract constraints from contextual
information for posts and extend traditional classifiers to model
these constraints such as LapRLS from Least Squares
in~\cite{belkin2006manifold} and LapSVM from Support Vector
Machines~\cite{sindhwani2005beyond}. To address the problem of data
classification with signed networks, we may need to understand the
structure of  positive and negative links in signed networks in
relation to attributes and labels of posts. For example, what are
the properties of posts from users with negative links in terms of
attributes and labels? If users have both positive and negative
links, what are the differences in terms of their posts? If users
with positive links are more likely to generate similar posts to
users with negative links, then the problem boils down to that of
classification with relative comparisons~\cite{schultz2004learning}.

\subsubsection{Data Clustering with Signed Networks}

Different from data classification, data clustering is unsupervised
learning, i.e., the label information ${\bf Y}$ is not available.
The problem of data clustering with signed networks is to find $f$
that identifies $k$ post clusters so that posts in the same cluster
are more similar to each other than to those in other clusters by
using information in $\{ {\bf X}, {\bf P}, {\bf A}\}$ and can be
formulated as follows:
\begin{align}
f: \{{\bf X}, {\bf P}, {\bf A}\} \rightarrow \{\mathcal{C}_1,\mathcal{C}_2,\ldots, \mathcal{C}_k\}
\end{align}
\noindent where $\mathcal{C}_i$ is the $i$-th clusters identified by a clustering function $f$.

{  By introducing the concept of pseudo-labels, unsupervised learning problems can be transformed into supervised
learning problems~\cite{masaeli2010transformation,cai2010unsupervised}. Hence an intuitive research direction is to transform
clustering with signed networks into classification with signed
networks with pseudo-labels.}
It is likely that posts from users with negative links may be from
different clusters and negative links may serve as additional
constraints when we cluster posts. Therefore, another possible
direction for data clustering with signed networks is to transform
data clustering algorithms with unsigned networks by considering
negative links as constraints and these constraints force posts from
users with negatives links to different clusters, which behaviors
similarly to traditional constraint clustering
problem~\cite{wagstaff2001constrained}. Recent research investigates
how to embed signed networks into a latent space where nodes sit
closer to their ``friends'' than their
``enemies''~\cite{cygan2012sitting,pardo2013embedding,kermarrec2014signed}.
Similarly we can develop algorithms to embed the combination of
signed networks and posts to learn representations for users and
posts simultaneously.

\section{Conclusions}

The availability of large-scale signed networks in social media has
encouraged  increasing attention on mining signed networks. Signed
networks are  unique in terms of basic concepts, principles and
properties of  specific computational tasks. This survey article
provides a comprehensive overview about mining signed networks in
social media. {  We first introduce basic concepts, principles and
properties of signed networks, including signed network representations (Section 2.1), properties of positive and negative links and social theories for signed networks}. Then,
we classify various tasks into node-oriented, link-oriented, and
application-oriented groups. Some of these tasks are well-studied,
whereas others need further investigation. For each group, we
review well-studied tasks with representative algorithms and also discuss some tasks that are not sufficiently studied
with formal definitions together with  promising research
directions.

In reviewing representative algorithms of well-studied tasks, for the methodology perspective, we notice that social theories such as balance theory and status theory are widely used in mining signed networks and we summarize three major ways in applying social theories in mining signed networks, i.e., feature engineering, constraint generating and objective defining.

\begin{itemize}
\item {\it Feature Engineering:} It helps extract features for computational models according to social theories. For example, in link prediction, triangle-based features are extracted based on balance theory to improve link predilection~\cite{leskovec2010predicting}, while triad features are extracted based on status theory to predict signs of links in~\cite{tang2012inferring}.
\item {\it Constraint Generating:} It generates constraints from social theories for computational models. Regularization is one of the most popular ways to implement constraint generating. For example, a regularization term is added to capture signed networks for recommendation based on generalized balance theory~\cite{forsati2014matrix}, and balance regularization is defined in~\cite{Tang-etal15a} to apply balance theory for negative link prediction.
\item {\it Objective Defining:} It uses social theories to define the objectives of the computational models. For example, In~\cite{amelio2013community}, based on balance theory, two objectives are developed for community detection, and balance theory and status theory are explicitly captured in the objective functions for sign prediction~\cite{yang2012friend}.
\end{itemize}

While from the technique perspective, we find that similar techniques such as random walk,  low-rank approximation and spectral clustering are adopted by various tasks of mining signed networks:
\begin{itemize}
\item Random Walk: Given a network and a starting node, we select one of its neighbors randomly, and move to the neighbor. Then we choose a neighbor of this node at random, and walk to it etc. The (random) sequence of nodes selected this way is a
random walk on the network~\cite{lovasz1993random}. The techniques of random walk are used in various tasks of mining signed networks such as node ranking~\cite{traag2010exponential} and community detection~\cite{yang2007community}.
\item Low-rank Approximation: Low-rank approximation aims to find a low-rank matrix such that the cost function, which measures the fit between the low-rank matrix and a given matrix, is optimized. It captures the low-rank structure of signed networks for link prediction~\cite{hsieh2012low} and it is one of the major techniques to build recommender systems with signed networks.
\item Spectral Clustering: Spectral clustering is derived from the graph partition problem, which aims to find a partition such that the cut (the number of links between two disjoint sets of nodes) is minimized. Spectral clustering is one of the most popular approaches for community detection~\cite{kunegis2010spectral}. Meanwhile, it can naturally generate vector representations for nodes thus it is also widely used in other tasks such as link prediction~\cite{chiang2013prediction}.
\end{itemize}

\section*{Acknowledgments}

This material is based upon work supported by, or in part by, the U.S. Army Research Office (ARO) under contract/grant
number 025071 and W911NF-15-1-0328, the Office of Naval Research(ONR) under grant number N000141410095, and the Army Research Laboratory and was accomplished under Cooperative Agreement Number W911NF-09-2-0053. The views and conclusions contained in this document are those of the authors and should not be interpreted as representing the official policies, either expressed or implied, of the Army Research Laboratory or the U.S. Government. The U.S. Government is authorized to reproduce and distribute reprints for Government purposes notwithstanding any copyright notation here on.

\bibliographystyle{ACM-Reference-Format-Journals}
\bibliography{signed}


\begin{thebibliography}{00}

%%% the Web address.

\ifx \showCODEN    \undefined \def \showCODEN     #1{\unskip}     \fi
\ifx \showDOI      \undefined \def \showDOI       #1{{\tt DOI:}\penalty0{#1}\ }
  \fi
\ifx \showISBNx    \undefined \def \showISBNx     #1{\unskip}     \fi
\ifx \showISBNxiii \undefined \def \showISBNxiii  #1{\unskip}     \fi
\ifx \showISSN     \undefined \def \showISSN      #1{\unskip}     \fi
\ifx \showLCCN     \undefined \def \showLCCN      #1{\unskip}     \fi
\ifx \shownote     \undefined \def \shownote      #1{#1}          \fi
\ifx \showarticletitle \undefined \def \showarticletitle #1{#1}   \fi
\ifx \showURL      \undefined \def \showURL       #1{#1}          \fi

\bibitem[\protect\citeauthoryear{Abbasi, Tang, and Liu}{Abbasi
  et~al\mbox{.}}{2014}]%
        {abbasi2014scalable}
{Mohammad~Ali Abbasi}, {Jiliang Tang}, {and} {Huan Liu}. 2014.
\newblock \showarticletitle{Scalable Learning of Users¡¯ Preferences Using
  Networked Data}. In {\em Proceedings of the 25th ACM conference on Hypertext
  and social media}. ACM, 4--12.
\newblock


\bibitem[\protect\citeauthoryear{Aggarwal and Subbian}{Aggarwal and
  Subbian}{2014}]%
        {aggarwal2014evolutionary}
{Charu Aggarwal} {and} {Karthik Subbian}. 2014.
\newblock \showarticletitle{Evolutionary network analysis: A survey}.
\newblock {\em ACM Computing Surveys (CSUR)\/} {47}, 1 (2014), 10.
\newblock


\bibitem[\protect\citeauthoryear{Aggarwal}{Aggarwal}{2011}]%
        {aggarwal2011introduction}
{Charu~C Aggarwal}. 2011.
\newblock {\em An introduction to social network data analytics}.
\newblock Springer.
\newblock


\bibitem[\protect\citeauthoryear{Agrawal, Garg, and Narayanam}{Agrawal
  et~al\mbox{.}}{2013}]%
        {agrawal2013link}
{Priyanka Agrawal}, {Vikas~K Garg}, {and} {Ramasuri Narayanam}. 2013.
\newblock \showarticletitle{Link label prediction in signed social networks}.
  In {\em Proceedings of the Twenty-Third international joint conference on
  Artificial Intelligence}. AAAI Press, 2591--2597.
\newblock


\bibitem[\protect\citeauthoryear{Ailon, Chen, and Huan}{Ailon
  et~al\mbox{.}}{2013}]%
        {ailon2013breaking}
{Nir Ailon}, {Yudong Chen}, {and} {Xu Huan}. 2013.
\newblock \showarticletitle{Breaking the small cluster barrier of graph
  clustering}.
\newblock {\em arXiv preprint arXiv:1302.4549\/} (2013).
\newblock


\bibitem[\protect\citeauthoryear{Amelio and Pizzuti}{Amelio and
  Pizzuti}{2013}]%
        {amelio2013community}
{Alessia Amelio} {and} {Clara Pizzuti}. 2013.
\newblock \showarticletitle{Community mining in signed networks: a
  multiobjective approach}. In {\em Proceedings of the 2013 IEEE/ACM
  International Conference on Advances in Social Networks Analysis and Mining}.
  ACM, 95--99.
\newblock


\bibitem[\protect\citeauthoryear{Anchuri and Magdon-Ismail}{Anchuri and
  Magdon-Ismail}{2012}]%
        {anchuri2012communities}
{Pranay Anchuri} {and} {Malik Magdon-Ismail}. 2012.
\newblock \showarticletitle{Communities and balance in signed networks: A
  spectral approach}. In {\em Advances in Social Networks Analysis and Mining
  (ASONAM), 2012 IEEE/ACM International Conference on}. IEEE, 235--242.
\newblock


\bibitem[\protect\citeauthoryear{Antal, Krapivsky, and Redner}{Antal
  et~al\mbox{.}}{2005}]%
        {antal2005dynamics}
{T Antal}, {PL Krapivsky}, {and} {S Redner}. 2005.
\newblock \showarticletitle{Dynamics of social balance on networks}.
\newblock {\em Physical Review E\/} {72}, 3 (2005), 036121.
\newblock


\bibitem[\protect\citeauthoryear{Axelrod and Bennett}{Axelrod and
  Bennett}{1993}]%
        {axelrod1993landscape}
{Robert Axelrod} {and} {D~Scott Bennett}. 1993.
\newblock \showarticletitle{A landscape theory of aggregation}.
\newblock {\em British journal of political science\/} {23}, 02 (1993),
  211--233.
\newblock


\bibitem[\protect\citeauthoryear{Bansal, Blum, and Chawla}{Bansal
  et~al\mbox{.}}{2004}]%
        {bansal2004correlation}
{Nikhil Bansal}, {Avrim Blum}, {and} {Shuchi Chawla}. 2004.
\newblock \showarticletitle{Correlation clustering}.
\newblock {\em Machine Learning\/} {56}, 1-3 (2004), 89--113.
\newblock


\bibitem[\protect\citeauthoryear{Beigi, Tang, and Liu}{Beigi
  et~al\mbox{.}}{2016}]%
        {beigi2016signed}
{Ghazaleh Beigi}, {Jiliang Tang}, {and} {Huan Liu}. 2016.
\newblock \showarticletitle{Signed Link Analysis in Social Media Networks}. In
  {\em Tenth International AAAI Conference on Web and Social Media}.
\newblock


\bibitem[\protect\citeauthoryear{Belkin and Niyogi}{Belkin and Niyogi}{2001}]%
        {belkin2001laplacian}
{Mikhail Belkin} {and} {Partha Niyogi}. 2001.
\newblock \showarticletitle{Laplacian Eigenmaps and Spectral Techniques for
  Embedding and Clustering.}. In {\em NIPS}, Vol.~14. 585--591.
\newblock


\bibitem[\protect\citeauthoryear{Belkin, Niyogi, and Sindhwani}{Belkin
  et~al\mbox{.}}{2006}]%
        {belkin2006manifold}
{Mikhail Belkin}, {Partha Niyogi}, {and} {Vikas Sindhwani}. 2006.
\newblock \showarticletitle{Manifold regularization: A geometric framework for
  learning from labeled and unlabeled examples}.
\newblock {\em The Journal of Machine Learning Research\/}  {7} (2006),
  2399--2434.
\newblock


\bibitem[\protect\citeauthoryear{Bhagat, Cormode, and Muthukrishnan}{Bhagat
  et~al\mbox{.}}{2011}]%
        {bhagat2011node}
{Smriti Bhagat}, {Graham Cormode}, {and} {S Muthukrishnan}. 2011.
\newblock \showarticletitle{Node classification in social networks}. In {\em
  Social network data analytics}. Springer, 115--148.
\newblock


\bibitem[\protect\citeauthoryear{Bilgic, Mihalkova, and Getoor}{Bilgic
  et~al\mbox{.}}{2010}]%
        {bilgic2010active}
{Mustafa Bilgic}, {Lilyana Mihalkova}, {and} {Lise Getoor}. 2010.
\newblock \showarticletitle{Active learning for networked data}. In {\em
  Proceedings of the 27th International Conference on Machine Learning
  (ICML-10)}. 79--86.
\newblock


\bibitem[\protect\citeauthoryear{Bogdanov, Larusso, and Singh}{Bogdanov
  et~al\mbox{.}}{2010}]%
        {bogdanov2010towards}
{Petko Bogdanov}, {Nicholas~D Larusso}, {and} {Ambuj Singh}. 2010.
\newblock \showarticletitle{Towards community discovery in signed collaborative
  interaction networks}. In {\em Data Mining Workshops (ICDMW), 2010 IEEE
  International Conference on}. IEEE, 288--295.
\newblock


\bibitem[\protect\citeauthoryear{Bonacich and Lloyd}{Bonacich and
  Lloyd}{2004}]%
        {bonacich2004calculating}
{Phillip Bonacich} {and} {Paulette Lloyd}. 2004.
\newblock \showarticletitle{Calculating status with negative relations}.
\newblock {\em Social Networks\/} {26}, 4 (2004), 331--338.
\newblock


\bibitem[\protect\citeauthoryear{Borgs, Chayes, Kalai, Malekian, and
  Tennenholtz}{Borgs et~al\mbox{.}}{2010}]%
        {borgs2010novel}
{Christian Borgs}, {Jennifer Chayes}, {Adam Kalai}, {Azarakhsh Malekian}, {and}
  {Moshe Tennenholtz}. 2010.
\newblock \showarticletitle{A Novel Approach to Propagating Distrust}.
\newblock {\em Internet and Network Economics\/} (2010), 87--105.
\newblock


\bibitem[\protect\citeauthoryear{Borzymek and Sydow}{Borzymek and
  Sydow}{2010}]%
        {borzymek2010trust}
{Piotr Borzymek} {and} {Marcin Sydow}. 2010.
\newblock \showarticletitle{Trust and distrust prediction in social network
  with combined graphical and review-based attributes}. In {\em Agent and
  Multi-Agent Systems: Technologies and Applications}. Springer, 122--131.
\newblock


\bibitem[\protect\citeauthoryear{Cai, Zhang, and He}{Cai et~al\mbox{.}}{2010}]%
        {cai2010unsupervised}
{Deng Cai}, {Chiyuan Zhang}, {and} {Xiaofei He}. 2010.
\newblock \showarticletitle{Unsupervised feature selection for multi-cluster
  data}. In {\em Proceedings of the 16th ACM SIGKDD international conference on
  Knowledge discovery and data mining}. ACM, 333--342.
\newblock


\bibitem[\protect\citeauthoryear{Cartwright and Gleason}{Cartwright and
  Gleason}{1966}]%
        {cartwright1966number}
{Dorwin Cartwright} {and} {Terry~C Gleason}. 1966.
\newblock \showarticletitle{The number of paths and cycles in a digraph}.
\newblock {\em Psychometrika\/} {31}, 2 (1966), 179--199.
\newblock


\bibitem[\protect\citeauthoryear{Cartwright and Harary}{Cartwright and
  Harary}{1956}]%
        {cartwright1956structural}
{Dorwin Cartwright} {and} {Frank Harary}. 1956.
\newblock \showarticletitle{Structural balance: a generalization of Heider's
  theory.}
\newblock {\em Psychological review\/} {63}, 5 (1956), 277.
\newblock


\bibitem[\protect\citeauthoryear{Cen, Gu, and Ji}{Cen et~al\mbox{.}}{2013}]%
        {cen2013sign}
{Yi Cen}, {Rentao Gu}, {and} {Yuefeng Ji}. 2013.
\newblock \showarticletitle{Sign Inference for Dynamic Signed Networks via
  Dictionary Learning}.
\newblock {\em Journal of Applied Mathematics\/}  {2013} (2013).
\newblock


\bibitem[\protect\citeauthoryear{Chen, Wan, Chung, and Sun}{Chen
  et~al\mbox{.}}{2013b}]%
        {chen2013effective}
{Chien~Chin Chen}, {Yu-Hao Wan}, {Meng-Chieh Chung}, {and} {Yu-Chun Sun}.
  2013b.
\newblock \showarticletitle{An effective recommendation method for cold start
  new users using trust and distrust networks}.
\newblock {\em Information Sciences\/}  {224} (2013), 19--36.
\newblock


\bibitem[\protect\citeauthoryear{Chen, Lakshmanan, and Castillo}{Chen
  et~al\mbox{.}}{2013a}]%
        {chen2013information}
{Wei Chen}, {Laks~VS Lakshmanan}, {and} {Carlos Castillo}. 2013a.
\newblock \showarticletitle{Information and Influence Propagation in Social
  Networks}.
\newblock {\em Synthesis Lectures on Data Management\/} {5}, 4 (2013), 1--177.
\newblock


\bibitem[\protect\citeauthoryear{Chen, Wang, and Yang}{Chen
  et~al\mbox{.}}{2009}]%
        {chen2009efficient}
{Wei Chen}, {Yajun Wang}, {and} {Siyu Yang}. 2009.
\newblock \showarticletitle{Efficient influence maximization in social
  networks}. In {\em Proceedings of the 15th ACM SIGKDD international
  conference on Knowledge discovery and data mining}. ACM, 199--208.
\newblock


\bibitem[\protect\citeauthoryear{Chen, Wang, and Yuan}{Chen
  et~al\mbox{.}}{2013}]%
        {chen2013overlapping}
{Yi Chen}, {Xiao-long Wang}, {and} {Bo Yuan}. 2013.
\newblock \showarticletitle{Overlapping community detection in signed
  networks}.
\newblock {\em arXiv preprint arXiv:1310.4023\/} (2013).
\newblock


\bibitem[\protect\citeauthoryear{Chiang, Hsieh, Natarajan, Tewari, and
  Dhillon}{Chiang et~al\mbox{.}}{2013}]%
        {chiang2013prediction}
{Kai-Yang Chiang}, {Cho-Jui Hsieh}, {Nagarajan Natarajan}, {Ambuj Tewari},
  {and} {Inderjit~S Dhillon}. 2013.
\newblock \showarticletitle{Prediction and Clustering in Signed Networks: A
  Local to Global Perspective}.
\newblock {\em arXiv preprint arXiv:1302.5145\/} (2013).
\newblock


\bibitem[\protect\citeauthoryear{Chiang, Natarajan, Tewari, and Dhillon}{Chiang
  et~al\mbox{.}}{2011}]%
        {chiang2011exploiting}
{Kai-Yang Chiang}, {Nagarajan Natarajan}, {Ambuj Tewari}, {and} {Inderjit~S
  Dhillon}. 2011.
\newblock \showarticletitle{Exploiting longer cycles for link prediction in
  signed networks}. In {\em Proceedings of the 20th ACM international
  conference on Information and knowledge management}. ACM, 1157--1162.
\newblock


\bibitem[\protect\citeauthoryear{Chiang, Whang, and Dhillon}{Chiang
  et~al\mbox{.}}{2012}]%
        {chiang2012scalable}
{Kai-Yang Chiang}, {Joyce~Jiyoung Whang}, {and} {Inderjit~S Dhillon}. 2012.
\newblock \showarticletitle{Scalable clustering of signed networks using
  balance normalized cut}. In {\em Proceedings of the 21st ACM international
  conference on Information and knowledge management}. ACM, 615--624.
\newblock


\bibitem[\protect\citeauthoryear{Cho}{Cho}{2006}]%
        {cho2006mechanism}
{Jinsook Cho}. 2006.
\newblock \showarticletitle{The mechanism of trust and distrust formation and
  their relational outcomes}.
\newblock {\em Journal of retailing\/} {82}, 1 (2006), 25--35.
\newblock


\bibitem[\protect\citeauthoryear{Chung, Tsiatas, and Xu}{Chung
  et~al\mbox{.}}{2013}]%
        {chung2013dirichlet}
{Fan Chung}, {Alexander Tsiatas}, {and} {Wensong Xu}. 2013.
\newblock \showarticletitle{Dirichlet PageRank and ranking algorithms based on
  trust and distrust}.
\newblock {\em Internet Mathematics\/} {9}, 1 (2013), 113--134.
\newblock


\bibitem[\protect\citeauthoryear{Clifford and Sudbury}{Clifford and
  Sudbury}{1973}]%
        {clifford1973model}
{Peter Clifford} {and} {Aidan Sudbury}. 1973.
\newblock \showarticletitle{A model for spatial conflict}.
\newblock {\em Biometrika\/} {60}, 3 (1973), 581--588.
\newblock


\bibitem[\protect\citeauthoryear{Cohn and Chang}{Cohn and Chang}{2000}]%
        {cohn2000learning}
{David Cohn} {and} {Huan Chang}. 2000.
\newblock \showarticletitle{Learning to probabilistically identify
  authoritative documents}. In {\em ICML}. Citeseer, 167--174.
\newblock


\bibitem[\protect\citeauthoryear{Coleman}{Coleman}{1988}]%
        {coleman1988social}
{James~S Coleman}. 1988.
\newblock \showarticletitle{Social capital in the creation of human capital}.
\newblock {\em American journal of sociology\/} (1988), S95--S120.
\newblock


\bibitem[\protect\citeauthoryear{Cygan, Pilipczuk, Pilipczuk, and
  Wojtaszczyk}{Cygan et~al\mbox{.}}{2012}]%
        {cygan2012sitting}
{Marek Cygan}, {Marcin Pilipczuk}, {Micha{\l} Pilipczuk}, {and} {Jakub~Onufry
  Wojtaszczyk}. 2012.
\newblock \showarticletitle{Sitting closer to friends than enemies, revisited}.
  In {\em Mathematical Foundations of Computer Science 2012}. Springer,
  296--307.
\newblock


\bibitem[\protect\citeauthoryear{Davis}{Davis}{1967}]%
        {davis1967clustering}
{James~A Davis}. 1967.
\newblock \showarticletitle{Clustering and structural balance in graphs.}
\newblock {\em Human relations\/} (1967).
\newblock


\bibitem[\protect\citeauthoryear{De~Cock and Da~Silva}{De~Cock and
  Da~Silva}{2006}]%
        {de2006many}
{Martine De~Cock} {and} {Paulo~Pinheiro Da~Silva}. 2006.
\newblock \showarticletitle{A many valued representation and propagation of
  trust and distrust}. In {\em Fuzzy Logic and Applications}. Springer,
  114--120.
\newblock


\bibitem[\protect\citeauthoryear{de~Kerchove et~al\mbox{.}}{de~Kerchove
  et~al\mbox{.}}{2009}]%
        {de2009ranking}
{Cristobald de Kerchove} {and} {others}. 2009.
\newblock {\em Ranking Large Networks: Leadership, Optimization and Distrust}.
\newblock Ph.D. Dissertation. UCL.
\newblock


\bibitem[\protect\citeauthoryear{De~Kerchove and Van~Dooren}{De~Kerchove and
  Van~Dooren}{2008}]%
        {de2008pagetrust}
{Cristobald De~Kerchove} {and} {Paul Van~Dooren}. 2008.
\newblock \showarticletitle{The PageTrust algorithm: How to rank web pages when
  negative links are allowed?}. In {\em SDM}. SIAM, 346--352.
\newblock


\bibitem[\protect\citeauthoryear{Doreian and Mrvar}{Doreian and Mrvar}{1996}]%
        {doreian1996partitioning}
{Patrick Doreian} {and} {Andrej Mrvar}. 1996.
\newblock \showarticletitle{A partitioning approach to structural balance}.
\newblock {\em Social networks\/} {18}, 2 (1996), 149--168.
\newblock


\bibitem[\protect\citeauthoryear{DuBois, Golbeck, and Srinivasan}{DuBois
  et~al\mbox{.}}{2011}]%
        {dubois2011predicting}
{Thomas DuBois}, {Jennifer Golbeck}, {and} {Aravind Srinivasan}. 2011.
\newblock \showarticletitle{Predicting trust and distrust in social networks}.
  In {\em Privacy, security, risk and trust (passat), 2011 ieee third
  international conference on and 2011 ieee third international conference on
  social computing (socialcom)}. IEEE, 418--424.
\newblock


\bibitem[\protect\citeauthoryear{Facchetti, Iacono, and Altafini}{Facchetti
  et~al\mbox{.}}{2011}]%
        {facchetti2011computing}
{Giuseppe Facchetti}, {Giovanni Iacono}, {and} {Claudio Altafini}. 2011.
\newblock \showarticletitle{Computing global structural balance in large-scale
  signed social networks}.
\newblock {\em Proceedings of the National Academy of Sciences\/} {108}, 52
  (2011), 20953--20958.
\newblock


\bibitem[\protect\citeauthoryear{Fan, Wang, Li, Li, and Jiang}{Fan
  et~al\mbox{.}}{2012}]%
        {fan2012analysis}
{Pengyi Fan}, {Hui Wang}, {Pei Li}, {Wei Li}, {and} {Zhihong Jiang}. 2012.
\newblock \showarticletitle{Analysis of opinion spreading in homogeneous
  networks with signed relationships}.
\newblock {\em Journal of Statistical Mechanics: Theory and Experiment\/}
  {2012}, 08 (2012), P08003.
\newblock


\bibitem[\protect\citeauthoryear{Forsati, Mahdavi, Shamsfard, and
  Sarwat}{Forsati et~al\mbox{.}}{2014}]%
        {forsati2014matrix}
{Rana Forsati}, {Mehrdad Mahdavi}, {Mehrnoush Shamsfard}, {and} {Mohamed
  Sarwat}. 2014.
\newblock \showarticletitle{Matrix Factorization with Explicit Trust and
  Distrust Side Information for Improved Social Recommendation}.
\newblock {\em ACM Transactions on Information Systems (TOIS)\/} {32}, 4
  (2014), 17.
\newblock


\bibitem[\protect\citeauthoryear{Frank and Harary}{Frank and Harary}{1980}]%
        {frank1980balance}
{Ove Frank} {and} {Frank Harary}. 1980.
\newblock \showarticletitle{Balance in stochastic signed graphs}.
\newblock {\em Social Networks\/} {2}, 2 (1980), 155--163.
\newblock


\bibitem[\protect\citeauthoryear{Getoor and Diehl}{Getoor and Diehl}{2005}]%
        {getoor2005link}
{Lise Getoor} {and} {Christopher~P Diehl}. 2005.
\newblock \showarticletitle{Link mining: a survey}.
\newblock {\em ACM SIGKDD Explorations Newsletter\/} {7}, 2 (2005), 3--12.
\newblock


\bibitem[\protect\citeauthoryear{Goldenberg, Libai, and Muller}{Goldenberg
  et~al\mbox{.}}{2001a}]%
        {goldenberg2001talk}
{Jacob Goldenberg}, {Barak Libai}, {and} {Eitan Muller}. 2001a.
\newblock \showarticletitle{Talk of the network: A complex systems look at the
  underlying process of word-of-mouth}.
\newblock {\em Marketing letters\/} {12}, 3 (2001), 211--223.
\newblock


\bibitem[\protect\citeauthoryear{Goldenberg, Libai, and Muller}{Goldenberg
  et~al\mbox{.}}{2001b}]%
        {goldenberg2001using}
{Jacob Goldenberg}, {Barak Libai}, {and} {Eitan Muller}. 2001b.
\newblock \showarticletitle{Using complex systems analysis to advance marketing
  theory development: Modeling heterogeneity effects on new product growth
  through stochastic cellular automata}.
\newblock {\em Academy of Marketing Science Review\/} {9}, 3 (2001), 1--18.
\newblock


\bibitem[\protect\citeauthoryear{G{\'o}mez, Jensen, and Arenas}{G{\'o}mez
  et~al\mbox{.}}{2009}]%
        {gomez2009analysis}
{Sergio G{\'o}mez}, {Pablo Jensen}, {and} {Alex Arenas}. 2009.
\newblock \showarticletitle{Analysis of community structure in networks of
  correlated data}.
\newblock {\em Physical Review E\/} {80}, 1 (2009), 016114.
\newblock


\bibitem[\protect\citeauthoryear{Granovetter}{Granovetter}{1978}]%
        {granovetter1978threshold}
{Mark Granovetter}. 1978.
\newblock \showarticletitle{Threshold models of collective behavior}.
\newblock {\em American journal of sociology\/} (1978), 1420--1443.
\newblock


\bibitem[\protect\citeauthoryear{Guha, Kumar, Raghavan, and Tomkins}{Guha
  et~al\mbox{.}}{2004}]%
        {guha2004propagation}
{Ramanthan Guha}, {Ravi Kumar}, {Prabhakar Raghavan}, {and} {Andrew Tomkins}.
  2004.
\newblock \showarticletitle{Propagation of trust and distrust}. In {\em
  Proceedings of the 13th international conference on World Wide Web}. ACM,
  403--412.
\newblock


\bibitem[\protect\citeauthoryear{Harary}{Harary}{1959}]%
        {harary1959measurement}
{Frank Harary}. 1959.
\newblock \showarticletitle{On the measurement of structural balance}.
\newblock {\em Behavioral Science\/} {4}, 4 (1959), 316--323.
\newblock


\bibitem[\protect\citeauthoryear{Harary et~al\mbox{.}}{Harary
  et~al\mbox{.}}{1953}]%
        {harary1953notion}
{Frank Harary} {and} {others}. 1953.
\newblock \showarticletitle{On the notion of balance of a signed graph.}
\newblock {\em The Michigan Mathematical Journal\/} {2}, 2 (1953), 143--146.
\newblock


\bibitem[\protect\citeauthoryear{Harary and Kabell}{Harary and Kabell}{1980}]%
        {harary1980simple}
{Frank Harary} {and} {Jerald~A Kabell}. 1980.
\newblock \showarticletitle{A simple algorithm to detect balance in signed
  graphs}.
\newblock {\em Mathematical Social Sciences\/} {1}, 1 (1980), 131--136.
\newblock


\bibitem[\protect\citeauthoryear{Harary and Kommel}{Harary and Kommel}{1979}]%
        {harary1979matrix}
{Frank Harary} {and} {Helene~J Kommel}. 1979.
\newblock \showarticletitle{Matrix measures for transitivity and balance*}.
\newblock {\em Journal of Mathematical Sociology\/} {6}, 2 (1979), 199--210.
\newblock


\bibitem[\protect\citeauthoryear{Hardin}{Hardin}{2004}]%
        {hardin2004distrust}
{Russell Hardin}. 2004.
\newblock \showarticletitle{Distrust: Manifestations and management}.
\newblock {\em Distrust\/}  {8} (2004), 3--33.
\newblock


\bibitem[\protect\citeauthoryear{Hassan, Abu-Jbara, and Radev}{Hassan
  et~al\mbox{.}}{2012a}]%
        {hassan2012detecting}
{Ahmed Hassan}, {Amjad Abu-Jbara}, {and} {Dragomir Radev}. 2012a.
\newblock \showarticletitle{Detecting subgroups in online discussions by
  modeling positive and negative relations among participants}. In {\em
  Proceedings of the 2012 Joint Conference on Empirical Methods in Natural
  Language Processing and Computational Natural Language Learning}. Association
  for Computational Linguistics, 59--70.
\newblock


\bibitem[\protect\citeauthoryear{Hassan, Abu-Jbara, and Radev}{Hassan
  et~al\mbox{.}}{2012b}]%
        {hassan2012extracting}
{Ahmed Hassan}, {Amjad Abu-Jbara}, {and} {Dragomir Radev}. 2012b.
\newblock \showarticletitle{Extracting signed social networks from text}. In
  {\em Workshop Proceedings of TextGraphs-7 on Graph-based Methods for Natural
  Language Processing}. Association for Computational Linguistics, 6--14.
\newblock


\bibitem[\protect\citeauthoryear{Haveliwala}{Haveliwala}{2002}]%
        {haveliwala2002topic}
{Taher~H Haveliwala}. 2002.
\newblock \showarticletitle{Topic-sensitive pagerank}. In {\em Proceedings of
  the 11th international conference on World Wide Web}. ACM, 517--526.
\newblock


\bibitem[\protect\citeauthoryear{Heider}{Heider}{1946}]%
        {heider1946attitudes}
{Fritz Heider}. 1946.
\newblock \showarticletitle{Attitudes and cognitive organization}.
\newblock {\em The Journal of psychology\/} {21}, 1 (1946), 107--112.
\newblock


\bibitem[\protect\citeauthoryear{Henley, Horsfall, and De~Soto}{Henley
  et~al\mbox{.}}{1969}]%
        {henley1969goodness}
{Nancy~M Henley}, {Robert~B Horsfall}, {and} {Clinton~B De~Soto}. 1969.
\newblock \showarticletitle{Goodness of figure and social structure.}
\newblock {\em Psychological Review\/} {76}, 2 (1969), 194.
\newblock


\bibitem[\protect\citeauthoryear{Hou}{Hou}{2005}]%
        {hou2005bounds}
{Yao~Ping Hou}. 2005.
\newblock \showarticletitle{Bounds for the least Laplacian eigenvalue of a
  signed graph}.
\newblock {\em Acta Mathematica Sinica\/} {21}, 4 (2005), 955--960.
\newblock


\bibitem[\protect\citeauthoryear{Hsieh, Chiang, and Dhillon}{Hsieh
  et~al\mbox{.}}{2012}]%
        {hsieh2012low}
{Cho-Jui Hsieh}, {Kai-Yang Chiang}, {and} {Inderjit~S Dhillon}. 2012.
\newblock \showarticletitle{Low rank modeling of signed networks}. In {\em
  Proceedings of the 18th ACM SIGKDD international conference on Knowledge
  discovery and data mining}. ACM, 507--515.
\newblock


\bibitem[\protect\citeauthoryear{Hu, Tang, Tang, and Liu}{Hu
  et~al\mbox{.}}{2013}]%
        {hu2013exploiting}
{Xia Hu}, {Lei Tang}, {Jiliang Tang}, {and} {Huan Liu}. 2013.
\newblock \showarticletitle{Exploiting social relations for sentiment analysis
  in microblogging}. In {\em Proceedings of the sixth ACM international
  conference on Web search and data mining}. ACM, 537--546.
\newblock


\bibitem[\protect\citeauthoryear{Huberman, Romero, and Wu}{Huberman
  et~al\mbox{.}}{2008}]%
        {huberman2008social}
{Bernardo~A Huberman}, {Daniel~M Romero}, {and} {Fang Wu}. 2008.
\newblock \showarticletitle{Social networks that matter: Twitter under the
  microscope}.
\newblock {\em arXiv preprint arXiv:0812.1045\/} (2008).
\newblock


\bibitem[\protect\citeauthoryear{Javari and Jalili}{Javari and Jalili}{2014}]%
        {javari2014cluster}
{Amin Javari} {and} {Mahdi Jalili}. 2014.
\newblock \showarticletitle{Cluster-Based Collaborative Filtering for Sign
  Prediction in Social Networks with Positive and Negative Links}.
\newblock {\em ACM Transactions on Intelligent Systems and Technology (TIST)\/}
  {5}, 2 (2014), 24.
\newblock


\bibitem[\protect\citeauthoryear{Jiang}{Jiang}{2015}]%
        {jiang2015stochastic}
{Jonathan~Q Jiang}. 2015.
\newblock \showarticletitle{Stochastic Blockmodel and Exploratory Analysis in
  Signed Networks}.
\newblock {\em arXiv preprint arXiv:1501.00594\/} (2015).
\newblock


\bibitem[\protect\citeauthoryear{Kamvar, Schlosser, and Garcia-Molina}{Kamvar
  et~al\mbox{.}}{2003}]%
        {kamvar2003eigentrust}
{Sepandar~D Kamvar}, {Mario~T Schlosser}, {and} {Hector Garcia-Molina}. 2003.
\newblock \showarticletitle{The eigentrust algorithm for reputation management
  in p2p networks}. In {\em Proceedings of the 12th international conference on
  World Wide Web}. ACM, 640--651.
\newblock


\bibitem[\protect\citeauthoryear{Kempe, Kleinberg, and Tardos}{Kempe
  et~al\mbox{.}}{2003}]%
        {kempe2003maximizing}
{David Kempe}, {Jon Kleinberg}, {and} {{\'E}va Tardos}. 2003.
\newblock \showarticletitle{Maximizing the spread of influence through a social
  network}. In {\em Proceedings of the ninth ACM SIGKDD international
  conference on Knowledge discovery and data mining}. ACM, 137--146.
\newblock


\bibitem[\protect\citeauthoryear{Kermarrec and Thraves}{Kermarrec and
  Thraves}{2014}]%
        {kermarrec2014signed}
{Anne-Marie Kermarrec} {and} {Christopher Thraves}. 2014.
\newblock \showarticletitle{Signed graph embedding: when everybody can sit
  closer to friends than enemies}.
\newblock {\em arXiv preprint arXiv:1405.5023\/} (2014).
\newblock


\bibitem[\protect\citeauthoryear{Kleinberg}{Kleinberg}{1999}]%
        {kleinberg1999authoritative}
{Jon~M Kleinberg}. 1999.
\newblock \showarticletitle{Authoritative sources in a hyperlinked
  environment}.
\newblock {\em Journal of the ACM (JACM)\/} {46}, 5 (1999), 604--632.
\newblock


\bibitem[\protect\citeauthoryear{Knapskog}{Knapskog}{1998}]%
        {knapskog1998metric}
{SJ Knapskog}. 1998.
\newblock \showarticletitle{A metric for trusted systems}. In {\em Proceedings
  of the 21st National Security Conference}. Citeseer, 16--29.
\newblock


\bibitem[\protect\citeauthoryear{Knoke and Yang}{Knoke and Yang}{2008}]%
        {knoke2008social}
{David Knoke} {and} {Song Yang}. 2008.
\newblock {\em Social network analysis}. Vol. 154.
\newblock Sage.
\newblock


\bibitem[\protect\citeauthoryear{Kong and Yang}{Kong and Yang}{2011}]%
        {kong2011improvement}
{Ling-Qi Kong} {and} {Meng-Long Yang}. 2011.
\newblock \showarticletitle{Improvement of clustering algorithm FEC for signed
  networks}.
\newblock {\em Journal of Computer Applications\/} {31}, 5 (2011), 1395--1399.
\newblock


\bibitem[\protect\citeauthoryear{Kunegis}{Kunegis}{2014}]%
        {kunegis2014applications}
{J{\'e}r{\^o}me Kunegis}. 2014.
\newblock \showarticletitle{Applications of Structural Balance in Signed Social
  Networks}.
\newblock {\em arXiv preprint arXiv:1402.6865\/} (2014).
\newblock


\bibitem[\protect\citeauthoryear{Kunegis, Lommatzsch, and Bauckhage}{Kunegis
  et~al\mbox{.}}{2009}]%
        {kunegis2009slashdot}
{J{\'e}r{\^o}me Kunegis}, {Andreas Lommatzsch}, {and} {Christian Bauckhage}.
  2009.
\newblock \showarticletitle{The slashdot zoo: mining a social network with
  negative edges}. In {\em Proceedings of the 18th international conference on
  World wide web}. ACM, 741--750.
\newblock


\bibitem[\protect\citeauthoryear{Kunegis, Preusse, and Schwagereit}{Kunegis
  et~al\mbox{.}}{2013}]%
        {kunegis2013added}
{J{\'e}r{\^o}me Kunegis}, {Julia Preusse}, {and} {Felix Schwagereit}. 2013.
\newblock \showarticletitle{What is the added value of negative links in online
  social networks?}. In {\em Proceedings of the 22nd international conference
  on World Wide Web}. International World Wide Web Conferences Steering
  Committee, 727--736.
\newblock


\bibitem[\protect\citeauthoryear{Kunegis, Schmidt, Lommatzsch, Lerner, De~Luca,
  and Albayrak}{Kunegis et~al\mbox{.}}{2010}]%
        {kunegis2010spectral}
{J{\'e}r{\^o}me Kunegis}, {Stephan Schmidt}, {Andreas Lommatzsch}, {J{\"u}rgen
  Lerner}, {Ernesto~William De~Luca}, {and} {Sahin Albayrak}. 2010.
\newblock \showarticletitle{Spectral analysis of signed graphs for clustering,
  prediction and visualization.}. In {\em SDM}, Vol.~10. SIAM, 559--559.
\newblock


\bibitem[\protect\citeauthoryear{Lecun, Yoshua, and Geoffrey}{Lecun
  et~al\mbox{.}}{2015}]%
        {yann2015}
{Yann Lecun}, {Bengio Yoshua}, {and} {Hinton Geoffrey}. 2015.
\newblock \showarticletitle{Deep Learning}.
\newblock {\em Nature\/} (2015).
\newblock


\bibitem[\protect\citeauthoryear{Leskovec, Huttenlocher, and
  Kleinberg}{Leskovec et~al\mbox{.}}{2010a}]%
        {leskovec2010predicting}
{Jure Leskovec}, {Daniel Huttenlocher}, {and} {Jon Kleinberg}. 2010a.
\newblock \showarticletitle{Predicting positive and negative links in online
  social networks}. In {\em Proceedings of the 19th international conference on
  World wide web}. ACM, 641--650.
\newblock


\bibitem[\protect\citeauthoryear{Leskovec, Huttenlocher, and
  Kleinberg}{Leskovec et~al\mbox{.}}{2010b}]%
        {leskovec2010signed}
{Jure Leskovec}, {Daniel Huttenlocher}, {and} {Jon Kleinberg}. 2010b.
\newblock \showarticletitle{Signed networks in social media}. In {\em
  Proceedings of the SIGCHI Conference on Human Factors in Computing Systems}.
  ACM, 1361--1370.
\newblock


\bibitem[\protect\citeauthoryear{Li, Xu, Chakraborty, Gupta, Sycara, and Li}{Li
  et~al\mbox{.}}{2014b}]%
        {li2014polarity}
{Dong Li}, {Zhi-Ming Xu}, {Nilanjan Chakraborty}, {Anika Gupta}, {Katia
  Sycara}, {and} {Sheng Li}. 2014b.
\newblock \showarticletitle{Polarity Related Influence Maximization in Signed
  Social Networks}.
\newblock {\em PloS one\/} {9}, 7 (2014), e102199.
\newblock


\bibitem[\protect\citeauthoryear{Li, Fan, Li, Wang, and Pan}{Li
  et~al\mbox{.}}{2013}]%
        {li2013opinion}
{Wei Li}, {Pengyi Fan}, {Pei Li}, {Hui Wang}, {and} {Yiguang Pan}. 2013.
\newblock \showarticletitle{An Opinion Spreading Model in Signed Networks}.
\newblock {\em Modern Physics Letters B\/} {27}, 12 (2013).
\newblock


\bibitem[\protect\citeauthoryear{Li, Chen, Wang, and Zhang}{Li
  et~al\mbox{.}}{2013}]%
        {li2013influence}
{Yanhua Li}, {Wei Chen}, {Yajun Wang}, {and} {Zhi-Li Zhang}. 2013.
\newblock \showarticletitle{Influence diffusion dynamics and influence
  maximization in social networks with friend and foe relationships}. In {\em
  Proceedings of the sixth ACM international conference on Web search and data
  mining}. ACM, 657--666.
\newblock


\bibitem[\protect\citeauthoryear{Li, Chen, Wang, and Zhang}{Li
  et~al\mbox{.}}{2014}]%
        {li2014voter}
{Yanhua Li}, {Wei Chen}, {Yajun Wang}, {and} {Zhi-Li Zhang}. 2014.
\newblock \showarticletitle{Voter Model on Signed Social Networks}.
\newblock {\em Internet Mathematics\/} just-accepted (2014).
\newblock


\bibitem[\protect\citeauthoryear{Li, Liu, and Liu}{Li et~al\mbox{.}}{2014a}]%
        {li2014comparative}
{Yadong Li}, {Jing Liu}, {and} {Chenlong Liu}. 2014a.
\newblock \showarticletitle{A comparative analysis of evolutionary and memetic
  algorithms for community detection from signed social networks}.
\newblock {\em Soft Computing\/} {18}, 2 (2014), 329--348.
\newblock


\bibitem[\protect\citeauthoryear{Liben-Nowell and Kleinberg}{Liben-Nowell and
  Kleinberg}{2007}]%
        {liben2007link}
{David Liben-Nowell} {and} {Jon Kleinberg}. 2007.
\newblock \showarticletitle{The link-prediction problem for social networks}.
\newblock {\em Journal of the American society for information science and
  technology\/} {58}, 7 (2007), 1019--1031.
\newblock


\bibitem[\protect\citeauthoryear{Long, Zhang, Wu, and Yu}{Long
  et~al\mbox{.}}{2006}]%
        {long2006spectral}
{Bo Long}, {Zhongfei~Mark Zhang}, {Xiaoyun Wu}, {and} {Philip~S Yu}. 2006.
\newblock \showarticletitle{Spectral clustering for multi-type relational
  data}. In {\em Proceedings of the 23rd international conference on Machine
  learning}. ACM, 585--592.
\newblock


\bibitem[\protect\citeauthoryear{Lov{\'a}sz}{Lov{\'a}sz}{1993}]%
        {lovasz1993random}
{L{\'a}szl{\'o} Lov{\'a}sz}. 1993.
\newblock \showarticletitle{Random walks on graphs: A survey}.
\newblock {\em Combinatorics, Paul erdos is eighty\/} {2}, 1 (1993), 1--46.
\newblock


\bibitem[\protect\citeauthoryear{Ma, Lyu, and King}{Ma et~al\mbox{.}}{2009}]%
        {ma2009learning}
{Hao Ma}, {Michael~R Lyu}, {and} {Irwin King}. 2009.
\newblock \showarticletitle{Learning to recommend with trust and distrust
  relationships}. In {\em Proceedings of the third ACM conference on
  Recommender systems}. ACM, 189--196.
\newblock


\bibitem[\protect\citeauthoryear{Ma, Zhou, Liu, Lyu, and King}{Ma
  et~al\mbox{.}}{2011}]%
        {ma2011recommender}
{Hao Ma}, {Dengyong Zhou}, {Chao Liu}, {Michael~R Lyu}, {and} {Irwin King}.
  2011.
\newblock \showarticletitle{Recommender systems with social regularization}. In
  {\em Proceedings of the fourth ACM international conference on Web search and
  data mining}. ACM, 287--296.
\newblock


\bibitem[\protect\citeauthoryear{Marsden and Friedkin}{Marsden and
  Friedkin}{1993}]%
        {marsden1993network}
{Peter~V Marsden} {and} {Noah~E Friedkin}. 1993.
\newblock \showarticletitle{Network studies of social influence}.
\newblock {\em Sociological Methods \& Research\/} {22}, 1 (1993), 127--151.
\newblock


\bibitem[\protect\citeauthoryear{Marvel, Kleinberg, Kleinberg, and
  Strogatz}{Marvel et~al\mbox{.}}{2011}]%
        {marvel2011continuous}
{Seth~A Marvel}, {Jon Kleinberg}, {Robert~D Kleinberg}, {and} {Steven~H
  Strogatz}. 2011.
\newblock \showarticletitle{Continuous-time model of structural balance}.
\newblock {\em Proceedings of the National Academy of Sciences\/} {108}, 5
  (2011), 1771--1776.
\newblock


\bibitem[\protect\citeauthoryear{Masaeli, Dy, and Fung}{Masaeli
  et~al\mbox{.}}{2010}]%
        {masaeli2010transformation}
{Mahdokht Masaeli}, {Jennifer~G Dy}, {and} {Glenn~M Fung}. 2010.
\newblock \showarticletitle{From transformation-based dimensionality reduction
  to feature selection}. In {\em Proceedings of the 27th International
  Conference on Machine Learning (ICML-10)}. 751--758.
\newblock


\bibitem[\protect\citeauthoryear{Massa and Avesani}{Massa and Avesani}{2005}]%
        {massa2005controversial}
{Paolo Massa} {and} {Paolo Avesani}. 2005.
\newblock \showarticletitle{Controversial users demand local trust metrics: An
  experimental study on epinions. com community}. In {\em Proceedings of the
  National Conference on artificial Intelligence}, Vol.~20. Menlo Park, CA;
  Cambridge, MA; London; AAAI Press; MIT Press; 1999, 121.
\newblock


\bibitem[\protect\citeauthoryear{May and Lloyd}{May and Lloyd}{2001}]%
        {may2001infection}
{Robert~M May} {and} {Alun~L Lloyd}. 2001.
\newblock \showarticletitle{Infection dynamics on scale-free networks}.
\newblock {\em Physical Review E\/} {64}, 6 (2001), 066112.
\newblock


\bibitem[\protect\citeauthoryear{McPherson, Smith-Lovin, and Cook}{McPherson
  et~al\mbox{.}}{2001}]%
        {mcpherson2001birds}
{Miller McPherson}, {Lynn Smith-Lovin}, {and} {James~M Cook}. 2001.
\newblock \showarticletitle{Birds of a feather: Homophily in social networks}.
\newblock {\em Annual review of sociology\/} (2001), 415--444.
\newblock


\bibitem[\protect\citeauthoryear{Mishra and Bhattacharya}{Mishra and
  Bhattacharya}{2011}]%
        {mishra2011finding}
{Abhinav Mishra} {and} {Arnab Bhattacharya}. 2011.
\newblock \showarticletitle{Finding the bias and prestige of nodes in networks
  based on trust scores}. In {\em Proceedings of the 20th international
  conference on World wide web}. ACM, 567--576.
\newblock


\bibitem[\protect\citeauthoryear{Moore}{Moore}{1978}]%
        {moore1978international}
{Michael Moore}. 1978.
\newblock \showarticletitle{An international application of Heider's balance
  theory}.
\newblock {\em European Journal of Social Psychology\/} {8}, 3 (1978),
  401--405.
\newblock


\bibitem[\protect\citeauthoryear{Moore}{Moore}{1979}]%
        {moore1979structural}
{Michael Moore}. 1979.
\newblock \showarticletitle{Structural balance and international relations}.
\newblock {\em European Journal of Social Psychology\/} {9}, 3 (1979),
  323--326.
\newblock


\bibitem[\protect\citeauthoryear{Nalluri}{Nalluri}{2014}]%
        {nalluriutility}
{Uma Nalluri}. 2014.
\newblock \showarticletitle{Utility of Distrust in Online Recommender Systems}.
\newblock {\em Capstone Project Report\/} (2014).
\newblock


\bibitem[\protect\citeauthoryear{Newman and Girvan}{Newman and Girvan}{2004}]%
        {newman2004finding}
{Mark~EJ Newman} {and} {Michelle Girvan}. 2004.
\newblock \showarticletitle{Finding and evaluating community structure in
  networks}.
\newblock {\em Physical review E\/} {69}, 2 (2004), 026113.
\newblock


\bibitem[\protect\citeauthoryear{Page, Brin, Motwani, and Winograd}{Page
  et~al\mbox{.}}{1999}]%
        {page1999pagerank}
{Lawrence Page}, {Sergey Brin}, {Rajeev Motwani}, {and} {Terry Winograd}. 1999.
\newblock \showarticletitle{The PageRank citation ranking: Bringing order to
  the web.}
\newblock  (1999).
\newblock


\bibitem[\protect\citeauthoryear{Papadopoulos, Kompatsiaris, Vakali, and
  Spyridonos}{Papadopoulos et~al\mbox{.}}{2012}]%
        {papadopoulos2012community}
{Symeon Papadopoulos}, {Yiannis Kompatsiaris}, {Athena Vakali}, {and}
  {Ploutarchos Spyridonos}. 2012.
\newblock \showarticletitle{Community detection in social media}.
\newblock {\em Data Mining and Knowledge Discovery\/} {24}, 3 (2012), 515--554.
\newblock


\bibitem[\protect\citeauthoryear{Papaoikonomou, Kardara, Tserpes, and
  Varvarigou}{Papaoikonomou et~al\mbox{.}}{2014}]%
        {papaoikonomou2014edge}
{Athanasios Papaoikonomou}, {Magdalini Kardara}, {Konstantinos Tserpes}, {and}
  {Dora Varvarigou}. 2014.
\newblock \showarticletitle{Edge Sign Prediction in Social Networks via
  Frequent Subgraph Discovery}.
\newblock {\em IEEE Internet Computing\/} (2014).
\newblock


\bibitem[\protect\citeauthoryear{Pardo, Soto, and Thraves}{Pardo
  et~al\mbox{.}}{2013}]%
        {pardo2013embedding}
{Eduardo~G Pardo}, {Mauricio Soto}, {and} {Christopher Thraves}. 2013.
\newblock \showarticletitle{Embedding signed graphs in the line}.
\newblock {\em Journal of Combinatorial Optimization\/} (2013), 1--21.
\newblock


\bibitem[\protect\citeauthoryear{Patidar, Agarwal, and Bharadwaj}{Patidar
  et~al\mbox{.}}{2012}]%
        {patidar2012predicting}
{Arti Patidar}, {Vinti Agarwal}, {and} {KK Bharadwaj}. 2012.
\newblock \showarticletitle{Predicting Friends and Foes in Signed Networks
  Using Inductive Inference and Social Balance Theory}. In {\em Proceedings of
  the 2012 International Conference on Advances in Social Networks Analysis and
  Mining (ASONAM 2012)}. IEEE Computer Society, 384--388.
\newblock


\bibitem[\protect\citeauthoryear{Pizzuti}{Pizzuti}{2009}]%
        {pizzuti2009multi}
{Clara Pizzuti}. 2009.
\newblock \showarticletitle{A multi-objective genetic algorithm for community
  detection in networks}. In {\em Tools with Artificial Intelligence, 2009.
  ICTAI'09. 21st International Conference on}. IEEE, 379--386.
\newblock


\bibitem[\protect\citeauthoryear{Qi and Davison}{Qi and Davison}{2009}]%
        {qi2009web}
{Xiaoguang Qi} {and} {Brian~D Davison}. 2009.
\newblock \showarticletitle{Web page classification: Features and algorithms}.
\newblock {\em ACM Computing Surveys (CSUR)\/} {41}, 2 (2009), 12.
\newblock


\bibitem[\protect\citeauthoryear{Radicchi, Vilone, and Meyer-Ortmanns}{Radicchi
  et~al\mbox{.}}{2007a}]%
        {radicchi2007universality}
{Filippo Radicchi}, {Daniele Vilone}, {and} {Hildegard Meyer-Ortmanns}. 2007a.
\newblock \showarticletitle{Universality class of triad dynamics on a
  triangular lattice}.
\newblock {\em Physical Review E\/} {75}, 2 (2007), 021118.
\newblock


\bibitem[\protect\citeauthoryear{Radicchi, Vilone, Yoon, and
  Meyer-Ortmanns}{Radicchi et~al\mbox{.}}{2007b}]%
        {radicchi2007social}
{Filippo Radicchi}, {Daniele Vilone}, {Sooeyon Yoon}, {and} {Hildegard
  Meyer-Ortmanns}. 2007b.
\newblock \showarticletitle{Social balance as a satisfiability problem of
  computer science}.
\newblock {\em Physical Review E\/} {75}, 2 (2007), 026106.
\newblock


\bibitem[\protect\citeauthoryear{Richardson, Agrawal, and Domingos}{Richardson
  et~al\mbox{.}}{2003}]%
        {richardson2003trust}
{Matthew Richardson}, {Rakesh Agrawal}, {and} {Pedro Domingos}. 2003.
\newblock \showarticletitle{Trust management for the semantic web}. In {\em The
  Semantic Web-ISWC 2003}. Springer, 351--368.
\newblock


\bibitem[\protect\citeauthoryear{Schelling}{Schelling}{2006}]%
        {schelling2006micromotives}
{Thomas~C Schelling}. 2006.
\newblock {\em Micromotives and macrobehavior}.
\newblock WW Norton \& Company.
\newblock


\bibitem[\protect\citeauthoryear{Schultz and Joachims}{Schultz and
  Joachims}{2004}]%
        {schultz2004learning}
{Matthew Schultz} {and} {Thorsten Joachims}. 2004.
\newblock \showarticletitle{Learning a distance metric from relative
  comparisons}.
\newblock {\em Advances in neural information processing systems (NIPS)\/}
  (2004), 41.
\newblock


\bibitem[\protect\citeauthoryear{Sen, Namata, Bilgic, Getoor, Galligher, and
  Eliassi-Rad}{Sen et~al\mbox{.}}{2008}]%
        {sen2008collective}
{Prithviraj Sen}, {Galileo Namata}, {Mustafa Bilgic}, {Lise Getoor}, {Brian
  Galligher}, {and} {Tina Eliassi-Rad}. 2008.
\newblock \showarticletitle{Collective classification in network data}.
\newblock {\em AI magazine\/} {29}, 3 (2008), 93.
\newblock


\bibitem[\protect\citeauthoryear{Shafaei and Jalili}{Shafaei and
  Jalili}{2014}]%
        {shafaei2014community}
{Mahsa Shafaei} {and} {Mahdi Jalili}. 2014.
\newblock \showarticletitle{Community Structure and Information Cascade in
  Signed Networks}.
\newblock {\em New Generation Computing\/} {32}, 3-4 (2014), 257--269.
\newblock


\bibitem[\protect\citeauthoryear{Shahriari and Jalili}{Shahriari and
  Jalili}{2014}]%
        {shahriari2014ranking}
{Moshen Shahriari} {and} {Mahdi Jalili}. 2014.
\newblock \showarticletitle{Ranking nodes in signed social networks}.
\newblock {\em Social Network Analysis and Mining\/} {4}, 1 (2014), 1--12.
\newblock


\bibitem[\protect\citeauthoryear{Sharma, Charls, and Singh}{Sharma
  et~al\mbox{.}}{2009}]%
        {sharma2009community}
{Tushar Sharma}, {Ankit Charls}, {and} {PramodKumar Singh}. 2009.
\newblock \showarticletitle{Community Mining in Signed Social Networks-An
  Automated Approach}.
\newblock {\em ICCEA09, Manila\/}  {9} (2009).
\newblock


\bibitem[\protect\citeauthoryear{Shen}{Shen}{2013}]%
        {shen2013community}
{Hua-Wei Shen}. 2013.
\newblock {\em Community Structure: An Introduction}.
\newblock Springer.
\newblock


\bibitem[\protect\citeauthoryear{Sindhwani, Niyogi, and Belkin}{Sindhwani
  et~al\mbox{.}}{2005}]%
        {sindhwani2005beyond}
{Vikas Sindhwani}, {Partha Niyogi}, {and} {Mikhail Belkin}. 2005.
\newblock \showarticletitle{Beyond the point cloud: from transductive to
  semi-supervised learning}. In {\em Proceedings of the 22nd international
  conference on Machine learning}. ACM, 824--831.
\newblock


\bibitem[\protect\citeauthoryear{Speriosu, Sudan, Upadhyay, and
  Baldridge}{Speriosu et~al\mbox{.}}{2011}]%
        {speriosu2011twitter}
{Michael Speriosu}, {Nikita Sudan}, {Sid Upadhyay}, {and} {Jason Baldridge}.
  2011.
\newblock \showarticletitle{Twitter polarity classification with label
  propagation over lexical links and the follower graph}. In {\em Proceedings
  of the First workshop on Unsupervised Learning in NLP}. Association for
  Computational Linguistics, 53--63.
\newblock


\bibitem[\protect\citeauthoryear{Srinivas and Deb}{Srinivas and Deb}{1994}]%
        {srinivas1994muiltiobjective}
{Nidamarthi Srinivas} {and} {Kalyanmoy Deb}. 1994.
\newblock \showarticletitle{Muiltiobjective optimization using nondominated
  sorting in genetic algorithms}.
\newblock {\em Evolutionary computation\/} {2}, 3 (1994), 221--248.
\newblock


\bibitem[\protect\citeauthoryear{Symeonidis and Mantas}{Symeonidis and
  Mantas}{2013}]%
        {symeonidis2013spectral}
{Panagiotis Symeonidis} {and} {Nikolaos Mantas}. 2013.
\newblock \showarticletitle{Spectral clustering for link prediction in social
  networks with positive and negative links}.
\newblock {\em Social Network Analysis and Mining\/} {3}, 4 (2013), 1433--1447.
\newblock


\bibitem[\protect\citeauthoryear{Symeonidis and Tiakas}{Symeonidis and
  Tiakas}{2013}]%
        {symeonidis2013transitive}
{Panagiotis Symeonidis} {and} {Eleftherios Tiakas}. 2013.
\newblock \showarticletitle{Transitive node similarity: predicting and
  recommending links in signed social networks}.
\newblock {\em World Wide Web\/} (2013), 1--34.
\newblock


\bibitem[\protect\citeauthoryear{Szell, Lambiotte, and Thurner}{Szell
  et~al\mbox{.}}{2010}]%
        {szell2010multirelational}
{Michael Szell}, {Renaud Lambiotte}, {and} {Stefan Thurner}. 2010.
\newblock \showarticletitle{Multirelational organization of large-scale social
  networks in an online world}.
\newblock {\em Proceedings of the National Academy of Sciences\/} {107}, 31
  (2010), 13636--13641.
\newblock


\bibitem[\protect\citeauthoryear{Tan, Lee, Tang, Jiang, Zhou, and Li}{Tan
  et~al\mbox{.}}{2011}]%
        {tan2011user}
{Chenhao Tan}, {Lillian Lee}, {Jie Tang}, {Long Jiang}, {Ming Zhou}, {and}
  {Ping Li}. 2011.
\newblock \showarticletitle{User-level sentiment analysis incorporating social
  networks}. In {\em Proceedings of the 17th ACM SIGKDD international
  conference on Knowledge discovery and data mining}. ACM, 1397--1405.
\newblock


\bibitem[\protect\citeauthoryear{Tang, Aggarwal, and Liu}{Tang
  et~al\mbox{.}}{2016a}]%
        {tang2015node}
{Jiliang Tang}, {Charu Aggarwal}, {and} {Huan Liu}. 2016a.
\newblock \showarticletitle{Node classification in signed social networks}. In
  {\em SIAM International Conference on Data Mining}.
\newblock


\bibitem[\protect\citeauthoryear{Tang, Aggarwal, and Liu}{Tang
  et~al\mbox{.}}{2016b}]%
        {tang2016recommendations}
{Jiliang Tang}, {Charu Aggarwal}, {and} {Huan Liu}. 2016b.
\newblock \showarticletitle{Recommendations in signed social networks}. In {\em
  Proceedings of the 25th International Conference on World Wide Web}.
  International World Wide Web Conferences Steering Committee, 31--40.
\newblock


\bibitem[\protect\citeauthoryear{Tang, Chang, Aggarwal, and Liu}{Tang
  et~al\mbox{.}}{2015}]%
        {Tang-etal15a}
{Jiliang Tang}, {Shiyu Chang}, {Charu Aggarwal}, {and} {Huan Liu}. 2015.
\newblock \showarticletitle{Negative Link Prediction in Social Media}. In {\em
  ACM International Conference on Web Search and Data Mining}.
\newblock


\bibitem[\protect\citeauthoryear{Tang, Hu, Chang, and Liu}{Tang
  et~al\mbox{.}}{2014b}]%
        {Tang-etal14d}
{Jiliang Tang}, {Xia Hu}, {Yi Chang}, {and} {Huan Liu}. 2014b.
\newblock \showarticletitle{Predictability of Distrust with Interaction Data}.
  In {\em ACM International Conference on Information and Knowledge
  Management}.
\newblock


\bibitem[\protect\citeauthoryear{Tang, Hu, and Liu}{Tang et~al\mbox{.}}{2013}]%
        {tang2013social}
{Jiliang Tang}, {Xia Hu}, {and} {Huan Liu}. 2013.
\newblock \showarticletitle{Social recommendation: a review}.
\newblock {\em Social Network Analysis and Mining\/} {3}, 4 (2013), 1113--1133.
\newblock


\bibitem[\protect\citeauthoryear{Tang, Hu, and Liu}{Tang
  et~al\mbox{.}}{2014a}]%
        {Tang-etal14b}
{Jiliang Tang}, {Xia Hu}, {and} {Huan Liu}. 2014a.
\newblock \showarticletitle{Is Distrust the Negation of Trust? The Value of
  Distrust in Social Media}. In {\em ACM Hypertext conference}.
\newblock


\bibitem[\protect\citeauthoryear{Tang and Liu}{Tang and Liu}{2012}]%
        {tang2012feature}
{Jiliang Tang} {and} {Huan Liu}. 2012.
\newblock \showarticletitle{Feature Selection with Linked Data in Social
  Media.}. In {\em SDM}. SIAM, 118--128.
\newblock


\bibitem[\protect\citeauthoryear{Tang, Lou, and Kleinberg}{Tang
  et~al\mbox{.}}{2012}]%
        {tang2012inferring}
{Jie Tang}, {Tiancheng Lou}, {and} {Jon Kleinberg}. 2012.
\newblock \showarticletitle{Inferring social ties across heterogenous
  networks}. In {\em Proceedings of the fifth ACM international conference on
  Web search and data mining}. ACM, 743--752.
\newblock


\bibitem[\protect\citeauthoryear{Tang, Tang, and Liu}{Tang
  et~al\mbox{.}}{2014}]%
        {tang2014recommendation}
{Jiliang Tang}, {Jie Tang}, {and} {Huan Liu}. 2014.
\newblock \showarticletitle{Recommendation in social media: recent advances and
  new frontiers}. In {\em Proceedings of the 20th ACM SIGKDD international
  conference on Knowledge discovery and data mining}. ACM, 1977--1977.
\newblock


\bibitem[\protect\citeauthoryear{Tang and Liu}{Tang and Liu}{2010}]%
        {tang2010community}
{Lei Tang} {and} {Huan Liu}. 2010.
\newblock \showarticletitle{Community detection and mining in social media}.
\newblock {\em Synthesis Lectures on Data Mining and Knowledge Discovery\/}
  {2}, 1 (2010), 1--137.
\newblock


\bibitem[\protect\citeauthoryear{Taylor}{Taylor}{1970}]%
        {taylor1970balance}
{Howard~Francis Taylor}. 1970.
\newblock {\em Balance in small groups}.
\newblock Van Nostrand Reinhold Company New
  York¡ªCincinnati¡ªToronto¡ªLondon¡ªMelbourne.
\newblock


\bibitem[\protect\citeauthoryear{Terzi and Winkler}{Terzi and Winkler}{2011}]%
        {terzi2011spectral}
{Evimaria Terzi} {and} {Marco Winkler}. 2011.
\newblock \showarticletitle{A spectral algorithm for computing social balance}.
  In {\em Algorithms and models for the web graph}. Springer, 1--13.
\newblock


\bibitem[\protect\citeauthoryear{Traag and Bruggeman}{Traag and
  Bruggeman}{2009}]%
        {traag2009community}
{VA Traag} {and} {Jeroen Bruggeman}. 2009.
\newblock \showarticletitle{Community detection in networks with positive and
  negative links}.
\newblock {\em Physical Review E\/} {80}, 3 (2009), 036115.
\newblock


\bibitem[\protect\citeauthoryear{Traag, Nesterov, and Van~Dooren}{Traag
  et~al\mbox{.}}{2010}]%
        {traag2010exponential}
{Vincent Traag}, {Yurii Nesterov}, {and} {Paul Van~Dooren}. 2010.
\newblock \showarticletitle{Exponential Ranking: Taking into Account Negative
  Links}.
\newblock {\em Social Informatics\/} (2010), 192--202.
\newblock


\bibitem[\protect\citeauthoryear{Victor, Cornelis, De~Cock, and Pinheiro~da
  Silva}{Victor et~al\mbox{.}}{2006}]%
        {victor2006towards}
{Patricia Victor}, {Chris Cornelis}, {Martine De~Cock}, {and} {P Pinheiro~da
  Silva}. 2006.
\newblock \showarticletitle{Towards a provenance-preserving trust model in
  agent networks}. In {\em WWW2006 Conference Proceedings, Special Interest
  Tracks, Posters and Workshops}.
\newblock


\bibitem[\protect\citeauthoryear{Victor, Cornelis, De~Cock, and
  Teredesai}{Victor et~al\mbox{.}}{2009}]%
        {victor2009trust}
{Patricia Victor}, {Chris Cornelis}, {Martine De~Cock}, {and} {Ankur
  Teredesai}. 2009.
\newblock \showarticletitle{Trust-and distrust-based recommendations for
  controversial reviews}. In {\em Web Science Conference (WebSci'09: Society
  On-Line)}.
\newblock


\bibitem[\protect\citeauthoryear{Victor, Verbiest, Cornelis, and Cock}{Victor
  et~al\mbox{.}}{2013}]%
        {victor2013enhancing}
{Patricia Victor}, {Nele Verbiest}, {Chris Cornelis}, {and} {Martine~De Cock}.
  2013.
\newblock \showarticletitle{Enhancing the trust-based recommendation process
  with explicit distrust}.
\newblock {\em ACM Transactions on the Web (TWEB)\/} {7}, 2 (2013), 6.
\newblock


\bibitem[\protect\citeauthoryear{Volz and Meyers}{Volz and Meyers}{2007}]%
        {volz2007susceptible}
{Erik Volz} {and} {Lauren~Ancel Meyers}. 2007.
\newblock \showarticletitle{Susceptible--infected--recovered epidemics in
  dynamic contact networks}.
\newblock {\em Proceedings of the Royal Society B: Biological Sciences\/}
  {274}, 1628 (2007), 2925--2934.
\newblock


\bibitem[\protect\citeauthoryear{Wagstaff, Cardie, Rogers, Schr{\"o}dl,
  et~al\mbox{.}}{Wagstaff et~al\mbox{.}}{2001}]%
        {wagstaff2001constrained}
{Kiri Wagstaff}, {Claire Cardie}, {Seth Rogers}, {Stefan Schr{\"o}dl}, {and}
  {others}. 2001.
\newblock \showarticletitle{Constrained k-means clustering with background
  knowledge}. In {\em ICML}, Vol.~1. 577--584.
\newblock


\bibitem[\protect\citeauthoryear{Wang, Ren, and Li}{Wang et~al\mbox{.}}{2014}]%
        {wang2014recognizing}
{Jun Wang}, {Fuji Ren}, {and} {Lei Li}. 2014.
\newblock \showarticletitle{Recognizing sentiment of relations between entities
  in text}.
\newblock {\em IEEJ Transactions on Electrical and Electronic Engineering\/}
  {9}, 6 (2014), 614--620.
\newblock


\bibitem[\protect\citeauthoryear{Wu, Aggarwal, and Sun}{Wu
  et~al\mbox{.}}{2016}]%
        {wu2016troll}
{Zhaoming Wu}, {Charu~C Aggarwal}, {and} {Jimeng Sun}. 2016.
\newblock \showarticletitle{The Troll-Trust Model for Ranking in Signed
  Networks}. In {\em Proceedings of the Ninth ACM International Conference on
  Web Search and Data Mining}. ACM, 447--456.
\newblock


\bibitem[\protect\citeauthoryear{Xiang, Neville, and Rogati}{Xiang
  et~al\mbox{.}}{2010}]%
        {xiang2010modeling}
{Rongjing Xiang}, {Jennifer Neville}, {and} {Monica Rogati}. 2010.
\newblock \showarticletitle{Modeling relationship strength in online social
  networks}. In {\em Proceedings of the 19th international conference on World
  wide web}. ACM, 981--990.
\newblock


\bibitem[\protect\citeauthoryear{Yang, Cheung, and Liu}{Yang
  et~al\mbox{.}}{2007}]%
        {yang2007community}
{Bo Yang}, {William~K Cheung}, {and} {Jiming Liu}. 2007.
\newblock \showarticletitle{Community mining from signed social networks}.
\newblock {\em Knowledge and Data Engineering, IEEE Transactions on\/} {19}, 10
  (2007), 1333--1348.
\newblock


\bibitem[\protect\citeauthoryear{Yang, Smola, Long, Zha, and Chang}{Yang
  et~al\mbox{.}}{2012}]%
        {yang2012friend}
{Shuang-Hong Yang}, {Alexander~J Smola}, {Bo Long}, {Hongyuan Zha}, {and} {Yi
  Chang}. 2012.
\newblock \showarticletitle{Friend or frenemy?: predicting signed ties in
  social networks}. In {\em Proceedings of the 35th international ACM SIGIR
  conference on Research and development in information retrieval}. ACM,
  555--564.
\newblock


\bibitem[\protect\citeauthoryear{Ye, Cheng, Zhu, and Chen}{Ye
  et~al\mbox{.}}{2013}]%
        {ye2013predicting}
{Jihang Ye}, {Hong Cheng}, {Zhe Zhu}, {and} {Minghua Chen}. 2013.
\newblock \showarticletitle{Predicting positive and negative links in signed
  social networks by transfer learning}. In {\em Proceedings of the 22nd
  international conference on World Wide Web}. International World Wide Web
  Conferences Steering Committee, 1477--1488.
\newblock


\bibitem[\protect\citeauthoryear{Yu and Xie}{Yu and Xie}{2014a}]%
        {yu2014learning}
{Xiaofeng Yu} {and} {Junqing Xie}. 2014a.
\newblock \showarticletitle{Learning Interactions for Social Prediction in
  Large-scale Networks}. In {\em Proceedings of the 23rd ACM International
  Conference on Conference on Information and Knowledge Management}. ACM,
  161--170.
\newblock


\bibitem[\protect\citeauthoryear{Yu and Xie}{Yu and Xie}{2014b}]%
        {yu2014modeling}
{Xiaofeng Yu} {and} {Jun~Qing Xie}. 2014b.
\newblock \showarticletitle{Modeling Mutual Influence Between Social Actions
  and Social Ties}.
\newblock  (2014).
\newblock


\bibitem[\protect\citeauthoryear{Zafarani, Abbasi, and Liu}{Zafarani
  et~al\mbox{.}}{2014}]%
        {zafarani2014social}
{Reza Zafarani}, {Mohammad~Ali Abbasi}, {and} {Huan Liu}. 2014.
\newblock {\em Social Media Mining: An Introduction}.
\newblock Cambridge University Press.
\newblock


\bibitem[\protect\citeauthoryear{Zhang, Zhan, He, Aggarwal, and Yu}{Zhang
  et~al\mbox{.}}{2016}]%
        {zhang2016trust}
{Jiawei Zhang}, {Qianyi Zhan}, {Lifang He}, {Charu Aggarwal}, {and} {Philip
  Yu}. 2016.
\newblock \showarticletitle{Trust Hole IdentiÞcation in Signed Networks}. In
  {\em The European Conference on Machine Learning and Principles and Practice
  of Knowledge Discovery}.
\newblock


\bibitem[\protect\citeauthoryear{Zhang, Jiang, Bao, and Zhang}{Zhang
  et~al\mbox{.}}{2013}]%
        {zhang2013characterization}
{Tongda Zhang}, {Haomiao Jiang}, {Zhouxiao Bao}, {and} {Yingfeng Zhang}. 2013.
\newblock \showarticletitle{Characterization and edge sign prediction in signed
  networks}.
\newblock {\em Journal of Industrial and Intelligent Information Vol\/} {1}, 1
  (2013).
\newblock


\bibitem[\protect\citeauthoryear{Zheleva and Getoor}{Zheleva and
  Getoor}{2009}]%
        {zheleva2009join}
{Elena Zheleva} {and} {Lise Getoor}. 2009.
\newblock \showarticletitle{To join or not to join: the illusion of privacy in
  social networks with mixed public and private user profiles}. In {\em
  Proceedings of the 18th international conference on World wide web}. ACM,
  531--540.
\newblock


\bibitem[\protect\citeauthoryear{Zheng and Skillicorn}{Zheng and
  Skillicorn}{2015}]%
        {zhengspectral}
{Q Zheng} {and} {DB Skillicorn}. 2015.
\newblock \showarticletitle{Spectral Embedding of Signed Networks}. In {\em
  SDM}. SIAM.
\newblock


\bibitem[\protect\citeauthoryear{Zheng, Zeng, and Wang}{Zheng
  et~al\mbox{.}}{2014}]%
        {zheng2014social}
{Xiaolong Zheng}, {Daniel Zeng}, {and} {Fei-Yue Wang}. 2014.
\newblock \showarticletitle{Social balance in signed networks}.
\newblock {\em Information Systems Frontiers\/} (2014), 1--19.
\newblock


\bibitem[\protect\citeauthoryear{Zhu, Yu, Chi, and Gong}{Zhu
  et~al\mbox{.}}{2007}]%
        {zhu2007combining}
{Shenghuo Zhu}, {Kai Yu}, {Yun Chi}, {and} {Yihong Gong}. 2007.
\newblock \showarticletitle{Combining content and link for classification using
  matrix factorization}. In {\em Proceedings of the 30th annual international
  ACM SIGIR conference on Research and development in information retrieval}.
  ACM, 487--494.
\newblock


\bibitem[\protect\citeauthoryear{Ziegler and Lausen}{Ziegler and
  Lausen}{2005}]%
        {ziegler2005propagation}
{Cai-Nicolas Ziegler} {and} {Georg Lausen}. 2005.
\newblock \showarticletitle{Propagation models for trust and distrust in social
  networks}.
\newblock {\em Information Systems Frontiers\/} {7}, 4-5 (2005), 337--358.
\newblock


\bibitem[\protect\citeauthoryear{Zolfaghar and Aghaie}{Zolfaghar and
  Aghaie}{2010}]%
        {zolfaghar2010mining}
{Kiyana Zolfaghar} {and} {Abdollah Aghaie}. 2010.
\newblock \showarticletitle{Mining trust and distrust relationships in social
  Web applications}. In {\em Intelligent Computer Communication and Processing
  (ICCP), 2010 IEEE International Conference on}. IEEE, 73--80.
\newblock


\end{thebibliography}

\end{document}